\documentclass[structabstract]{aa}

\usepackage{txfonts}
\usepackage{natbib}
\usepackage{amsbsy}
\usepackage{graphicx}

\bibpunct{(}{)}{;}{a}{}{,}     

\newcommand{\micron}{$\mu$m}

\newcommand{\el}[1]{\mathrm{#1}}

\newcommand{\ten}[1]{$10^{#1}$}
\newcommand{\scit}[2]{$#1\times10^{#2}$}
\newcommand{\scim}[2]{#1\times10^{#2}}

\newcommand{\ps}{s$^{-1}$}
\newcommand{\pcc}{cm$^{-3}$}
\newcommand{\eq}[1]{Eq.\ (\ref{eq:#1})}

\newcommand{\fig}[1]{Fig.\ \ref{fig:#1}}
\newcommand{\figg}[1]{Figure \ref{fig:#1}}
\newcommand{\tb}[1]{Table \ref{tb:#1}}

\newcommand{\pder}[2]{\frac{\partial #1}{\partial #2}}

\newcommand{\erf}{\mathrm{erf}}
\newcommand{\tacc}{t_\el{acc}}
\newcommand{\cs}{c_\el{s}}
\newcommand{\rcen}{R_\el{c}}
\newcommand{\w}{H$_2$O}

\newcommand{\fmh}{H$_2$CO}
\newcommand{\meoh}{CH$_3$OH}
\newcommand{\mf}{HCOOCH$_3$}

\defcitealias{shu77a}{S77}
\defcitealias{cassen81a}{CM81}
\defcitealias{terebey84a}{TSC84}
\defcitealias{young05a}{YE05}
\defcitealias{brinch08b}{BWH08}

\begin{document}

\title{The chemical history of molecules in circumstellar disks. I. Ices}

\author{
R. Visser \inst{1}
 \and E.F. van Dishoeck \inst{1,2}
 \and S.D. Doty \inst{3}
 \and C.P. Dullemond \inst{4}
}

\institute{Leiden Observatory, Leiden University, P.O. Box 9513, 2300 RA Leiden, The Netherlands \\
           \email{ruvisser@strw.leidenuniv.nl}
 \and Max-Planck-Institut f\"ur Extraterrestrische Physik, Giessenbachstrasse 1, 85748 Garching, Germany
 \and Department of Physics and Astronomy, Denison University, Granville, OH 43023, USA
 \and Max-Planck-Institut f\"{u}r Astronomie, Koenigstuhl 17, 69117 Heidelberg, Germany
}

\authorrunning{Visser et al.}

\date{Received $<$date$>$ / Accepted $<$date$>$}


\abstract
{Many chemical changes occur during the collapse of a molecular cloud to form a low-mass star and the surrounding disk. One-dimensional models have been used so far to analyse these chemical processes, but they cannot properly describe the incorporation of material into disks.} 
{The goal of this work is to understand how material changes chemically as it is transported from the cloud to the star and the disk. Of special interest is the chemical history of the material in the disk at the end of the collapse.} 
{A two-dimensional, semi-analytical model is presented that follows, for the first time, the chemical evolution from the pre-stellar core to the protostar and circumstellar disk. The model computes infall trajectories from any point in the cloud and tracks the radial and vertical motion of material in the viscously evolving disk. It includes a full time-dependent radiative transfer treatment of the dust temperature, which controls much of the chemistry. A small parameter grid is explored to understand the effects of the sound speed and the mass and rotation of the cloud. The freeze-out and evaporation of carbon monoxide (CO) and water (\w), as well as the potential for forming complex organic molecules in ices, are considered as important first steps to illustrate the full chemistry.} 
{Both species freeze out towards the centre before the collapse begins. Pure CO ice evaporates during the infall phase and re-adsorbs in those parts of the disk that cool below the CO desorption temperature of $\sim$18 K. \w{} remains solid almost everywhere during the infall and disk formation phases and evaporates within $\sim$10 AU of the star. Mixed CO-\w{} ices are important in keeping some solid CO above 18 K and in explaining the presence of CO in comets. Material that ends up in the planet- and comet-forming zones of the disk ($\sim$5--30 AU from the star) is predicted to spend enough time in a warm zone (several \ten{4} yr at a dust temperature of 20--40 K) during the collapse to form first-generation complex organic species on the grains. The dynamical timescales in the hot inner envelope (hot core or hot corino) are too short for abundant formation of second-generation molecules by high-temperature gas-phase chemistry.} 
{} 


\maketitle


\section{Introduction}
\label{sec:intro}
The formation of low-mass stars and their planetary systems is a complex event, spanning several orders of magnitude in temporal and spatial scales, and involving a wide variety of physical and chemical processes. Thanks to observations (see reviews by \citealt{difrancesco07a} and \citealt{white07a}), theory \citep[see review by][]{shu87a} and computer simulations (see reviews by \citealt{klein07a} and \citealt{dullemond07b}), the general picture of low-mass star formation is now understood. An instability in a cold molecular cloud leads to gravitational collapse. Rotation and magnetic fields cause a flattened density structure early on, which evolves into a circumstellar disk at later times. The protostar continues to accrete matter from the disk and the remnant envelope, while also expelling matter in a bipolar pattern. Grain growth in the disk eventually leads to the formation of planets, and as the remaining dust and gas disappear, a mature solar system emerges. While there has been ample discussion in the literature on the origin and evolution of grains in disks (see reviews by \citealt{natta07a} and \citealt{dominik07a}), little attention has so far been paid to the chemical history of the more volatile material in a two- or three-dimensional setting.

Chemical models are required to understand the observations and develop the simulations (see reviews by \citealt{ceccarelli07a}, \citealt{bergin07a} and \citealt{bergin07b}). The chemistry in pre-stellar cores is relatively easy to model, because the dynamics and the temperature structure are simpler before the protostar is formed than afterwards. A key result from the pre-stellar core models is the depletion of many carbon-bearing species towards the centre of the core \citep{bergin97b,lee04a}.

\citet{ceccarelli96a} modelled the chemistry in the collapse phase, and others have done so more recently \citep{rodgers03a,doty04a,lee04a,garrod06a,aikawa08a,garrod08a}. All of these models are one-dimensional, and thus necessarily ignore the circumstellar disk. As the protostar turns on and heats up the surrounding material, all models agree that frozen-out species return to the gas phase if the dust temperature surpasses their evaporation temperature. The higher temperatures can further drive a hot-core-like chemistry, and complex molecules may be formed if the infall timescales are long enough.

If the model is expanded into a second dimension and the disk is included, the system gains a large reservoir where infalling material from the cloud can be stored for a long time before accreting onto the star. This can lead to further chemical enrichment, especially in the warmer parts of the disk \citep{aikawa97a,aikawa99a,willacy00a,vanzadelhoff03a,rodgers03a,aikawa08a}. The interior of the disk is shielded from direct irradiation by the star, so it is colder than the disk's surface and the remnant cloud. Hence, molecules that evaporated as they fell in towards the star may freeze out again when they enter the disk. This was first shown quantitatively by \citet[hereafter BWH08]{brinch08b} using a two-dimensional hydrodynamical simulation.

In addition to observations of nearby star-forming regions, the comets in our own solar system provide a unique probe into the chemistry that takes place during star and planet formation. The bulk composition of the cometary nuclei is believed to be largely pristine, closely reflecting the composition of the pre-solar nebula \citep{bockelee04a}. However, large abundance variations have been observed between individual comets and these remain poorly understood \citep{kobayashi07a}. Two-dimensional chemical models may shed light on the cometary chemical diversity.

Two molecules of great astrophysical interest are carbon monoxide (CO) and water (\w). They are the main reservoirs of carbon and oxygen and control much of the chemistry. CO is an important precursor for more complex molecules; for example, solid CO can be hydrogenated to formaldehyde (\fmh) and methanol (\meoh) at low temperatures (\citealt{watanabe02a}; Fuchs et al. subm.). In turn, these two molecules form the basis of even larger organic species like methyl formate (\mf; \citealt{garrod06a}; \citealt{garrod08a}). The key role of \w{} in the formation of life on Earth and potentially elsewhere is evident. If the entire formation process of low-mass stars and their planets is to be understood, a thorough understanding of these two molecules is essential.

This paper is the first in a series aiming to model the chemical evolution from the pre-stellar core to the disk phase in two dimensions, using a simplified, semi-analytical approach for the dynamics of the collapsing envelope and the disk, but including detailed radiative transfer for the temperature structure. The model follows individual parcels of material as they fall in from the cloud into the disk. The gaseous and solid abundances of CO and \w{} are calculated for each infalling parcel to obtain global gas-ice profiles. The semi-analytical nature of the model allows for an easy exploration of physical parameters like the cloud's mass and rotation rate, or the effective sound speed. Tracing the temperature history of the infalling material provides a first clue into the formation of more complex species. The model also provides some insight into the origin of the chemical diversity in comets.

Section \ref{sec:model} contains a full description of the model. Results are presented in Section \ref{sec:stanres} and discussed in a broader astrophysical context in Section \ref{sec:disc}. Conclusions are drawn in Section \ref{sec:conc}.


\section{Model}
\label{sec:model}
The physical part of our two-dimensional axisymmetric model describes the collapse of an initially spherical, isothermal, slowly rotating cloud to form a star and circumstellar disk. The collapse dynamics are taken from \citet[hereafter S77]{shu77a}, including the effects of rotation as described by \citet[hereafter CM81]{cassen81a} and \citet[hereafter TSC84]{terebey84a}. Infalling material hits the equatorial plane inside the centrifugal radius to form a disk, whose further evolution is constrained by conservation of angular momentum \citep{lyndenbell74a}. Some properties of the star and the disk are adapted from \citet{adams86a} and \citet[hereafter YE05]{young05a}. Magnetic fields are not included in our model. They are unlikely to affect the chemistry directly and their main physical effect (causing a flattened density distribution; \citealt{galli93a}) is already accounted for by the rotation of the cloud.

Our model is an extension of the one used by \citet{dullemond06a} to study the crystallinity of dust in circumstellar disks. That model was purely one-dimensional; our model treats the disk more realistically as a two-dimensional structure.


\subsection{Envelope}
\label{subsec:env}
The cloud (or envelope) is taken to be a uniformly rotating singular isothermal sphere at the onset of collapse. It has a solid-body rotation rate $\Omega_0$ and an $r^{-2}$ density profile \citepalias{shu77a}:
\begin{equation}
\label{eq:sisrho}
\rho_0(r) = \frac{\cs^2}{2\pi Gr^2}\,,
\end{equation}
where $G$ is the gravitational constant and $\cs$ the effective sound speed. Throughout this work, $r$ is used for the spherical radius and $R$ for the cylindrical radius. Setting the outer radius at $r_\el{env}$, the total mass of the cloud is
\begin{equation}
\label{eq:sismass}
M_0 = \frac{2\cs^2r_\el{env}}{G}\,.
\end{equation}

After the collapse is triggered at the centre, an expansion wave (or collapse front) travels outwards at the sound speed \citepalias{shu77a,terebey84a}. Material inside the expansion wave falls in towards the centre to form a protostar. The infalling material is deflected towards the gravitational midplane by the cloud's rotation. It first hits the midplane inside the centrifugal radius (where gravity balances angular momentum; \citetalias{cassen81a}), resulting in the formation of a circumstellar disk (Section \ref{subsec:disk}).

The dynamics of a collapsing singular isothermal sphere were computed by \citetalias{shu77a} in terms of the non-dimensional variable $x=r/\cs t$, with $t$ the time after the onset of collapse. In this self-similar description, the head of the expansion wave is always at $x=1$. The density and radial velocity are given by the non-dimensional variables $\mathcal{A}$ and $v$, respectively. (\citetalias{shu77a} uses $\alpha$ for the density, but our model already uses that symbol for the viscosity in Section \ref{subsec:disk}.) These variables are dimensionalised through
\begin{equation}
\label{eq:shuden}
\rho(r,t) = \frac{\mathcal{A}(x)}{4\pi Gt^2}\,,
\end{equation}
\begin{equation}
\label{eq:shuvel}
u_r(r,t) = \cs v(x)\,.
\end{equation}
Values for $\mathcal{A}$ and $v$ are tabulated in \citetalias{shu77a}.

\citetalias{cassen81a} and \citetalias{terebey84a} analysed the effects of slow uniform rotation on the \citetalias{shu77a} collapse solution, with the former focussing on the flow onto the protostar and the disk and the latter on what happens further out in the envelope. In the axisymmetric \citetalias{terebey84a} description, the density and infall velocities depend on the time, $t$, the radius, $r$, and the polar angle, $\theta$:
\begin{equation}
\label{eq:tscden}
\rho(r,\theta,t) = \frac{\mathcal{A}(x,\theta,\tau)}{4\pi Gt^2}\,,
\end{equation}
\begin{equation}
\label{eq:tscrvel}
u_r(r,\theta,t) = \cs v(x,\theta,\tau)\,,
\end{equation}
where $\tau=\Omega_0t$ is the non-dimensional time. The polar velocity is given by
\begin{equation}
\label{eq:tscthvel}
u_\theta(r,\theta,t) = \cs w(x,\theta,\tau)\,.
\end{equation}
The differential equations from \citetalias{terebey84a} were solved numerically to obtain solutions for $\mathcal{A}$, $v$ and $w$.

The \citetalias{terebey84a} solution breaks down around $x=\tau^2$, so the \citetalias{cassen81a} solution is used inside of this point. A streamline through a point $(r,\theta)$ effectively originated at an angle $\theta_0$ in this description:
\begin{equation}
\label{eq:cmthetafull}
\frac{\cos\theta_0-\cos\theta}{\sin^2\theta_0\cos\theta_0} - \frac{\rcen}{r} = 0\,,
\end{equation}
where $\rcen$ is the centrifugal radius,
\begin{equation}
\label{eq:centr}
\rcen(t) = \frac{1}{16}\cs m_0^3t^3\Omega_0^2\,,
\end{equation}
with $m_0$ a numerical factor equal to 0.975. The \citetalias{cassen81a} radial and polar velocity are
\begin{equation}
\label{eq:cmrvel}
u_r(r,\theta,t) = -\sqrt{\frac{GM}{r}}\sqrt{1+\frac{\cos\theta}{\cos\theta_0}}\,,
\end{equation}
\begin{equation}
\label{eq:cmthvel}
u_\theta(r,\theta,t) = \sqrt{\frac{GM}{r}}\sqrt{1+\frac{\cos\theta}{\cos\theta_0}}\frac{\cos\theta_0-\cos\theta}{\sin\theta}\,,
\end{equation}
and the \citetalias{cassen81a} density is
\begin{equation}
\label{eq:cmden}
\rho(r,\theta,t) = -\frac{\dot{M}}{4\pi{}r^2u_r}\left[1+2\frac{\rcen}{r}P_2(\cos\theta_0)\right]^{-1}\,,
\end{equation}
where $P_2$ is the second-order Legendre polynomial and $\dot{M}=m_0\cs^3/G$ is the total accretion rate from the envelope onto the star and disk \citepalias{shu77a,terebey84a}. The primary accretion phase ends when the outer shell of the envelope reaches the star and disk. This point in time ($\tacc=M_0/\dot{M}$) is essentially the beginning of the T Tauri or Herbig Ae/Be phase, but it does not yet correspond to a typical T Tauri or Herbig Ae/Be object (see Section \ref{subsec:stanrestemp}).

The \citetalias{terebey84a} and \citetalias{cassen81a} solutions do not reproduce the cavities created by the star's bipolar outflow, so they have to be put in separately. Outflows have been observed in two shapes: conical and curved \citep{padgett99a}. Both can be characterized by the outflow opening angle, $\gamma$, which grows with the age of the object. \citet{arce06a} found a linear relationship in log-log space between the age of a sample of 17 young stellar objects and their outflow opening angles. Some explanations exist for the outflow widening in general, but it is not yet understood how $\gamma(t)$ depends on parameters like the initial cloud mass and the sound speed. It is likely that the angle depends on the relative age of the object rather than on the absolute age.

The purpose of our model is not to include a detailed description of the outflow cavity. Instead, the outflow is primarily included because of its effect on the temperature profiles \citep{whitney03a}. Its opening angle is based on the fit by \citet{arce06a} to their Fig.\ 5, but it is taken to depend on $t/\tacc$ rather than $t$ alone. The outflow is also kept smaller, which brings it closer to the \citeauthor{whitney03a} angles. Its shape is taken to be conical. With the resulting formula,
\begin{equation}
\label{eq:outflow}
\log\frac{\gamma(t)}{\deg} = 1.5 + 0.26\log\frac{t}{\tacc}\,,
\end{equation}
the opening angle is always 32$^\circ$ at $t=\tacc$. The numbers in \eq{outflow} are poorly constrained; however, the details of the outflow (both size and shape) do not affect the temperature profiles strongly, so this introduces no major errors in the chemistry results. The outflow cones are filled with a constant mass of $0.002M_0$ at a uniform density, which decreases to \ten{3}--\ten{4} \pcc{} at $\tacc$ depending on the model parameters. The outflow effectively removes about 1\% of the initial envelope mass.


\subsection{Disk}
\label{subsec:disk}
The rotation of the envelope causes the infalling material to be deflected towards the midplane, where it forms a circumstellar disk. The disk initially forms inside the centrifugal radius \citepalias{cassen81a}, but conservation of angular momentum quickly causes the disk to spread beyond this point. The evolution of the disk is governed by viscosity, for which our model uses the common $\alpha$ prescription \citep{shakura73a}. This gives the viscosity coefficient $\nu$ as
\begin{equation}
\label{eq:viscco}
\nu(R,t) = \alpha c_\el{s,d}H\,.
\end{equation}
The sound speed in the disk, $c_\el{s,d}=\sqrt{kT_\el{m}/\mu m_\el{p}}$ (with $k$ the Boltzmann constant, $m_\el{p}$ the proton mass and $\mu$ the mean molecular mass of 2.3 nuclei per hydrogen molecule), is different from the sound speed in the envelope, $\cs$, because the midplane temperature of the disk, $T_\el{m}$, varies as described in \citet{hueso05a}. The other variable from \eq{viscco} is the scale height:
\begin{equation}
\label{eq:scale}
H(R,t) = \frac{c_\el{s,d}}{\Omega_\el{k}}\,,
\end{equation}
where $\Omega_\el{k}$ is the Keplerian rotation rate:
\begin{equation}
\label{eq:kep}
\Omega_\el{k}(R,t) = \sqrt{\frac{GM_\ast}{R^3}}\,,
\end{equation}
with $M_\ast$ the stellar mass [\eq{mstar}]. The viscosity parameter $\alpha$ is kept constant at \ten{-2} \citep{dullemond07b,andrews07a}.

Solving the problem of advection and diffusion yields the radial velocities inside the disk \citep{dullemond06a,lyndenbell74a}:
\begin{equation}
\label{eq:veldisk}
u_R(R,t) = -\frac{3}{\Sigma\sqrt{R}}\pder{}{R}\left(\Sigma\nu\sqrt{R}\right)\,.
\end{equation}
The surface density evolves as
\begin{equation}
\label{eq:surfevol}
\pder{\Sigma(R,t)}{t} = -\frac{1}{R}\pder{}{R}(\Sigma Ru_R) + S\,,
\end{equation}
where the source function $S$ accounts for the infall of material from the envelope:
\begin{equation}
\label{eq:source}
S(R,t) = 2N\rho u_z\,,
\end{equation}
with $u_z$ the vertical component of the envelope velocity field [Eqs.\ (\ref{eq:tscrvel}), (\ref{eq:tscthvel}), (\ref{eq:cmrvel}) and (\ref{eq:cmthvel})]. The factor 2 accounts for the envelope accreting onto both sides of the disk and the normalization factor $N$ ensures that the overall accretion rate onto the star and the disk is always equal to $\dot{M}$. Both $\rho$ and $u_z$ in \eq{source} are to be computed at the disk-envelope boundary, which will be defined at the end of this section.

As noted by \citet{hueso05a}, the infalling envelope material enters the disk with a subkeplerian rotation rate, so, by conservation of angular momentum, it would tend to move a bit further inwards. Not taking this into account would artificially generate angular momentum, causing the disk to take longer to accrete onto the star. As a consequence the disk will, at any given point in time, have too high a mass and too large a radius. \citeauthor{hueso05a} solved this problem by modifying \eq{source} to place the material directly at the correct radius. However, this causes an undesirable discontinuity in the infall trajectories. Instead, our model adds a small extra component to \eq{veldisk} for $t<\tacc$:
\begin{equation}
\label{eq:veldisk2}
u_R(R,t) = -\frac{3}{\Sigma\sqrt{R}}\pder{}{R}\left(\Sigma\nu\sqrt{R}\right)\, - \eta_\el{r}\sqrt{\frac{GM}{R}}.
\end{equation}
The functional form of the extra term derives from the CM81 solution. A constant value of 0.002 for $\eta_\el{r}$ is found to reproduce very well the results of \citeauthor{hueso05a}. It also provides a good match with the disk masses from \citet{yorke99a}, \citetalias{young05a} and \citetalias{brinch08b}, whose models cover a wide range of initial conditions.

The disk's inner radius is determined by the evaporation of dust by the star (e.g. \citetalias{young05a}):
\begin{equation}
\label{eq:diskin}
R_\el{i}(t) = \sqrt{\frac{L_\ast}{4\pi\sigma T_\el{evap}^4}}\,,
\end{equation}
where $\sigma$ is the Stefan-Boltzmann constant. The dust evaporation temperature, $T_\el{evap}$, is set to 2000 K. Taking an alternative value of 1500 K has no effect on our results. The stellar luminosity, $L_\ast$, is discussed in Section \ref{subsec:star}. Inward transport of material at $R_\el{i}$ leads to accretion from the disk onto the star:
\begin{equation}
\label{eq:mdotds}
\dot{M}_{\el{d}\to\ast} = -2\pi{}R_\el{i}u_R\Sigma\,,
\end{equation}
with the radial velocity, $u_R$, and the surface density, $\Sigma$, taken at $R_\el{i}$. The disk gains mass from the envelope at a rate $\dot{M}_{\el{e}\to\el{d}}$, so the disk mass evolves as
\begin{equation}
\label{eq:mdisk}
M_\el{d}(t) = \int_0^t\left(\dot{M}_{\el{e}\to\el{d}}-\dot{M}_{\el{d}\to\ast}\right)\el{d}t'\,.
\end{equation}

Our model uses a Gaussian profile for the vertical structure of the disk \citep{shakura73a}:
\begin{equation}
\label{eq:rhoz}
\rho(R,z,t) = \rho_\el{c}\exp\left(-\frac{z^2}{2H^2}\right)\,,
\end{equation}
with $z$ the height above the midplane. The scale height comes from \eq{scale} and the midplane density is
\begin{equation}
\label{eq:rhom}
\rho_\el{c}(R,t) = \frac{\Sigma}{H\sqrt{2\pi}}\,.
\end{equation}
Along with the radial motion [\eq{veldisk2}, taken to be independent of $z$], material also moves vertically in the disk, as it must maintain the Gaussian profile at all times. To see this, consider a parcel of material that enters the disk at time $t$ at coordinates $R$ and $z$ into a column with scale height $H$ and surface density $\Sigma$. The column of material below the parcel is
\begin{equation}
\label{eq:velvert1}
\int_0^{z}\rho(R,\zeta,t)\el{d}\zeta = \frac{1}{2}\Sigma\erf\left(\frac{z}{H\sqrt{2}}\right)\,,
\end{equation}
where $\erf$ is the error function. At a later time $t'$, the entire column has moved to $R'$ and has a scale height $H'$ and a surface density $\Sigma'$. The same amount of material must still be below the parcel:
\begin{equation}
\label{eq:velvert2}
\frac{1}{2}\Sigma'\erf\left(\frac{z'}{H'\sqrt{2}}\right) = \frac{1}{2}\Sigma\erf\left(\frac{z}{H\sqrt{2}}\right)\,.
\end{equation}
Rearranging gives the new height of the parcel, $z'$:
\begin{equation}
\label{eq:velvert3}
z'(R',t') = H'\sqrt{2}\erf^{-1}\left[\frac{\Sigma}{\Sigma'}\erf\left(\frac{z}{H\sqrt{2}}\right)\right]\,,
\end{equation}
where $\erf^{-1}$ is the inverse of the error function. In the absence of mixing, our description leads to purely laminar flow.

\begin{figure}
\resizebox{\hsize}{!}{\includegraphics{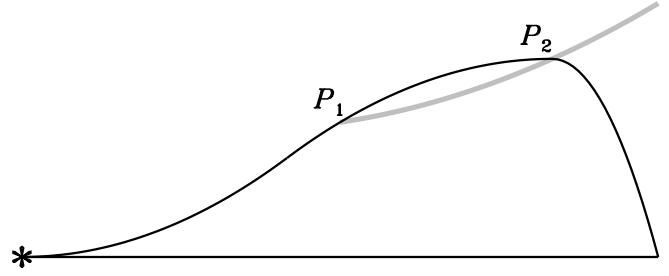}}
\caption{Schematic view of the disk-envelope boundary in the ($R,z$) plane. The black line indicates the surface where the density due to the disk equals that due to the envelope. The grey line is the infall trajectory that would lead to point $P_1$. However, it already intersects the disk at point $P_2$, so no accretion is possible at $P_1$. The disk-envelope boundary is therefore raised at $P_1$ until it can be reached freely by an infall trajectory.}
\label{fig:boundary}
\end{figure}

The location of the disk-envelope boundary [needed, for instance, in \eq{source}] is determined in two steps. First, the surface is identified where the density due to the disk [\eq{rhoz}] equals that due to the envelope [Eqs.\ (\ref{eq:tscden}) and (\ref{eq:cmden})]. In order for accretion to take place at a given point $P_1$ on the surface, it must be intersected by an infall trajectory. Due to the geometry of the surface, such a trajectory might also intersect the disk at a larger radius $P_2$ (\fig{boundary}). Material flowing in along that trajectory will accrete at $P_2$ instead of $P_1$. Hence, the second step in determining the disk-envelope boundary consists of raising the surface at ``obstructed points'' like $P_1$ to an altitude where accretion can take place. The source function is then computed at that altitude. Physically, this can be understood as follows: the region directly above the obstructed points becomes less dense than what it would be in the absence of a disk, because the disk also prevents material from reaching there. The lower density above the disk reduces the downward pressure, so the disk puffs up and the disk-envelope boundary moves to a higher altitude.

The infall trajectories in the vicinity of the disk are very shallow, so the bulk of the material accretes at the outer edge. Because the disk quickly spreads beyond the centrifugal radius, much of the accretion occurs far from the star. In contrast, accretion in one-dimensional collapse models occurs at or inside of $R_\el{c}$. Our results are consistent with the hydrodynamical work of \citetalias{brinch08b}, where most of the infalling material also hits the outer edge of a rather large disk. The large accretion radii lead to weaker accretion shocks than commonly assumed (Section \ref{subsec:shock}).


\subsection{Star}
\label{subsec:star}
{\setlength\arraycolsep{2pt}
The star gains material from the envelope and from the disk, so its mass evolves as
\begin{equation}
\label{eq:mstar}
M_\ast(t) = \int_0^t\left(\dot{M}_{\el{e}\to\ast}+\dot{M}_{\el{d}\to\ast}\right)\el{d}t'\,.
\end{equation}
The protostar does not come into existence immediately at the onset of collapse; it is preceeded by the first hydrostatic core \citep[FHC;][]{masunaga98a,boss95a}. Our model follows \citetalias{young05a} and takes a lifetime of \scit{2}{4} yr and a size of 5 AU for the FHC, independent of other parameters. After this stage, a rapid transition occurs from the large FHC to a protostar of a few $R_\odot$:
\begin{eqnarray}
\label{eq:rstar}
R_\ast & = & (5\,\el{AU})\left(1-\sqrt{\frac{t-20,000\,\el{yr}}{100\,\el{yr}}}\right)+R_\ast^\el{PS} \nonumber\\
& & 20,000 < t\,\el{(yr)} < 20,100\,,
\end{eqnarray}
where $R_\ast^\el{PS}$ (ranging from 2 to 5 $R_\odot$) is the protostellar radius from \citet{palla91a}. For $t>\scim{2.01}{4}$ yr, $R_\ast$ equals $R_\ast^\el{PS}$. Our results are not sensitive to the exact values used for the size and lifetime of the FHC or the duration of the FHC--protostar transition.
}

{\setlength\arraycolsep{2pt}
The star's luminosity, $L_\ast$, consists of two terms: the accretion luminosity, $L_{\ast,\el{acc}}$, dominant at early times, and the luminosity due to gravitational contraction and deuterium burning, $L_\el{phot}$. The accretion luminosity comes from \citet{adams86a}:
\begin{eqnarray}
\label{eq:l*acc}
L_{\ast,\el{acc}}(t) & = & L_0\bigg\{\frac{1}{6u_\ast}\left[6u_\ast-2+(2-5u_\ast)\sqrt{1-u_\ast}\right] + \nonumber\\
& & \frac{\eta_\ast}{2}\left[1-(1-\eta_\el{d})\mathcal{M}_\el{d}\right]\left[1-(1-\eta_\el{d})\sqrt{1-u_\ast}\right]\bigg\}\,,
\end{eqnarray}
where $L_0=GM\dot{M}/R_\ast$ (with $M$ the total accreted mass, i.e., $M=M_\ast+M_\el{d}$), $u_\ast=R_\ast/\rcen$, and
\begin{equation}
\label{eq:mmmd}
\mathcal{M}_\el{d} = \frac{1}{3}u_\ast^{1/3}\int_{u_\ast}^1\frac{\sqrt{1-u}}{u^{4/3}}\el{d}u\,.
\end{equation}
Analytical solutions exist for the asymptotic cases of $u_\ast\approx0$ and $u_\ast\approx1$. For intermediate values, the integral must be solved numerically. The efficiency parameters $\eta_\ast$ and $\eta_\el{d}$ in \eq{l*acc} have values of 0.5 and 0.75 for a 1 $M_\odot$ envelope \citepalias{young05a}. The photospheric luminosity is adopted from \citet{dantona94a}, using \citetalias{young05a}'s method of fitting and interpolating, including a time difference of 0.38 $\tacc$ (equal to the free-fall time) between the onset of $L_{\ast,\el{acc}}$ and $L_{\ast,\el{phot}}$ \citep{myers98a}. The sum of these two terms gives the total stellar luminosity:
\begin{equation}
\label{eq:lstar}
L_\ast(t) = L_{\ast,\el{acc}} + L_{\ast,\el{phot}}\,.
\end{equation}
}

\begin{figure}
\resizebox{\hsize}{!}{\includegraphics{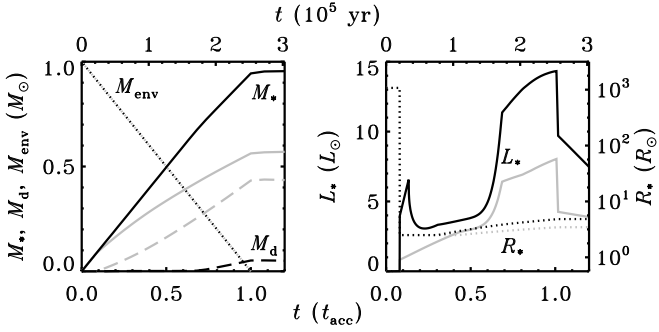}}
\caption{Evolution of the mass of the envelope, star and disk (left panel) and the luminosity (solid lines) and radius (dotted lines) of the star (right panel) for our standard model (black lines) and our reference model (grey lines).}
\label{fig:evol}
\end{figure}

\figg{evol} shows the evolution of the stellar mass, luminosity and radius for our standard case of $M_0=1.0$ $M_\odot$, $\cs=0.26$ km \ps{} and $\Omega_0=10^{-14}$ \ps{}, and our reference case of $M_0=1.0$ $M_\odot$, $\cs=0.26$ km \ps{} and $\Omega_0=10^{-13}$ \ps{} (Section \ref{subsec:params}). The transition from the FHC to the protostar at $t=\scim{2}{4}$ yr is clearly visible in the $R_\ast$ and $L_\ast$ profiles. At $t=\tacc$, there is no more accretion from the envelope onto the star, so the luminosity decreases sharply.

The masses of the disk and the envelope are also shown in \fig{evol}. Our disk mass of 0.43 $M_\odot$ at $t=\tacc$ in the reference case is in excellent agreement with the value of 0.4 $M_\odot$ found by \citetalias{brinch08b} for the same parameters.


\subsection{Temperature}
\label{subsec:temp}
The envelope starts out as an isothermal sphere at 10 K and it is heated up from the inside after the onset of collapse. Using the star as the only photon source, the dust temperature in the disk and envelope is computed with the axisymmetric three-dimensional radiative transfer code RADMC \citep{dullemond04a}. Because of the high densities throughout most of the system, the gas and dust are expected to be well coupled, and the gas temperature is set equal to the dust temperature. Cosmic-ray heating of the gas is included implicitly by setting a lower limit of 8 K in the dust radiative transfer results. As mentioned in Section \ref{subsec:env}, the presence of the outflow cones has some effect on the temperature profiles \citep{whitney03a}. This will be discussed further in Section \ref{subsec:stanrestemp}.


\subsection{Accretion shock}
\label{subsec:shock}
The infall of high-velocity envelope material into the low-velocity disk causes a J-type shock. The temperature right behind the shock front can be much higher than what it would be due to the stellar photons. \citet{neufeld94a} calculated in detail the relationship between the pre-shock velocities and densities ($u_\el{s}$ and $n_\el{s}$) and the maximum grain temperature reached after the shock ($T_\el{d,s}$). A simple formula, valid for $u_\el{s}<70$ km \ps, can be extracted from their Fig.\ 13:
\begin{equation}
\label{eq:shock}
T_\el{d,s} \approx (104\,\el{K})\left(\frac{n_\el{s}}{10^6\,\el{cm}^{-3}}\right)^{0.21}\left(\frac{u_\el{s}}{30\,\el{km}\,\el{s}^{-1}}\right)^p\left(\frac{a_\el{gr}}{0.1\,\mu\el{m}}\right)^{-0.20}\,,
\end{equation}
with $a_\el{gr}$ the grain radius. The exponent $p$ is 0.62 for $u_\el{s}<30$ km \ps{} and 1.0 otherwise.

\begin{figure}
\resizebox{\hsize}{!}{\includegraphics{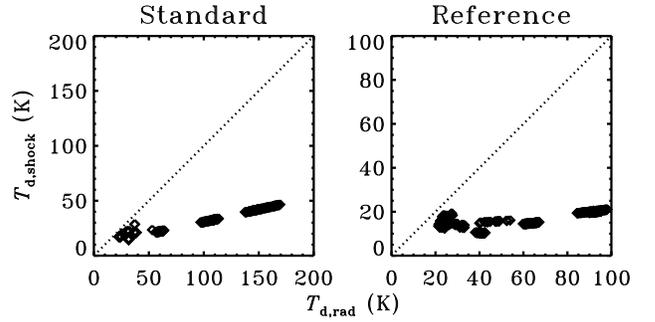}}
\caption{Dust temperature due to the accretion shock (vertical axis) and stellar radiation (horizontal axis) at the point of entry into the disk for 0.1-\micron{} grains in a sample of several hundred parcels in our standard (left) and reference (right) models. These parcels occupy positions from $R=1$ to 300 AU in the disk at $\tacc$. Shock heating is not important for the dust in most of these parcels, because the shock temperature is lower than the radiative heating temperature for all of them. Note the different scales between the two panels.}
\label{fig:shock}
\end{figure}

The pre-shock velocities and densities are highest at early times, when accretion occurs close to the star and all ices would evaporate anyway. Important for our purposes is the question whether the dust temperature due to the shock exceeds that due to stellar heating. If all grains have a radius of 0.1 \micron, as assumed in our model, this is not the case for any of the material in the disk at $\tacc$ for either our standard or our reference model (\fig{shock}; cf. \citealt{simonelli97a}).

In reality, the dust spans a range of sizes, extending down to a radius of about 0.005 \micron. Small grains are heated more easily; 0.005-\micron{} dust reaches a shock temperature almost twice as high as does 0.1-\micron{} dust [\eq{shock}]. This is enough for the shock temperature to exceed the radiative heating temperature in part of the sample in \fig{shock}. However, this has no effect on the CO and \w{} gas-ice ratios. In those parcels where shock heating becomes important for small grains, the temperature from radiative heating lies already above the CO evaporation temperature of about 18 K and the shock temperature remains below 60 K, which is not enough for \w{} to evaporate. Hence, shock heating is not included in our model.

\w{} may also be removed from the grain surfaces in the accretion shock through sputtering \citep{tielens94a,jones94a}. The material that makes up the disk at the end of the collapse in our standard model experiences a shock of at most 8 km \ps. At that velocity, \element[+]{He}, the most important ion for sputtering, carries an energy of 1.3 eV. However, a minimum of 2.2 eV is required to remove \w{} \citep{bohdansky80a}, so sputtering is unimportant for our purposes.


Some of the material in our model is heated to more than 100 K during the collapse (\fig{stanhist}) or experiences a shock strong enough to induce sputtering. This material normally ends up in the star before the end of the collapse, but mixing may keep some of it in the disk. The possible consequences are discussed briefly in Section \ref{subsec:comets}.




\subsection{Model parameters}
\label{subsec:params}
The standard set of parameters for our model corresponds to Case J from \citet{yorke99a}, except that the solid-body rotation rate is reduced from \ten{-13} to \ten{-14} \ps{} to produce a more realistic disk mass of 0.05 $M_\odot$, consistent with observations \citep[e.g.][]{andrews07b,andrews07a}. The envelope has an initial mass of 1.0 $M_\odot$ and a radius of 6700 AU, and the effective sound speed is 0.26 km \ps{}.

The original Case J (with $\Omega_0=10^{-13}$ \ps), which was also used in \citetalias{brinch08b}, is used here as a reference model to enable a direct quantitative comparison of the results with an independent method. This case results in a much higher disk mass of 0.43 $M_\odot$. Although such high disk masses are not excluded by observations and theoretical arguments \citep{hartmann06a}, they are considered less representative of typical young stellar objects than the disks of lower mass.

The parameters $M_0$, $\cs$ and $\Omega_0$ are changed in one direction each to create a $2^3$ parameter grid. The two values for $\Omega_0$, \ten{-14} and \ten{-13} \ps, cover the range of rotation rates observed by \citet{goodman93a}. The other variations are chosen for their opposite effect: a lower sound speed gives a more massive disk, and a lower initial mass gives a less massive disk. The full model is run for each set of parameters to analyse how the chemistry can vary between different objects. The parameter grid is summarised in \tb{pgrid}. Our standard set is Case 3 and our reference set is Case 7.

\begin{table}
\caption{Summary of the parameter grid used in our model.$^{\mathrm{a}}$}
\label{tb:pgrid}
\centering
\begin{tabular}{cccccccc}
\hline\hline
Case$^{\mathrm{b}}$ & $\Omega_0$ & $\cs$    & $M_0$       & $\tacc$      & $\tau_\el{ads}$ & $M_\el{d}$  \\
                  & (\ps)      & (km \ps) & ($M_\odot$) & (\ten{5} yr) & (\ten{5} yr) & ($M_\odot$) \\
\hline
1 \phantom{(std)} & \ten{-14} & 0.19 & 1.0 & 6.3 & 14.4           & 0.22\phantom{0} \\
2 \phantom{(std)} & \ten{-14} & 0.19 & 0.5 & 3.2 & \phantom{0}3.6 & 0.08\phantom{0} \\
3 (std)           & \ten{-14} & 0.26 & 1.0 & 2.5 & \phantom{0}2.3 & 0.05\phantom{0} \\
4 \phantom{(std)} & \ten{-14} & 0.26 & 0.5 & 1.3 & \phantom{0}0.6 & 0.001 \\
5 \phantom{(std)} & \ten{-13} & 0.19 & 1.0 & 6.3 & 14.4           & 0.59\phantom{0} \\
6 \phantom{(std)} & \ten{-13} & 0.19 & 0.5 & 3.2 & \phantom{0}3.6 & 0.25\phantom{0} \\
7 (ref)           & \ten{-13} & 0.26 & 1.0 & 2.5 & \phantom{0}2.3 & 0.43\phantom{0} \\
8 \phantom{(std)} & \ten{-13} & 0.26 & 0.5 & 1.3 & \phantom{0}0.6 & 0.16\phantom{0} \\
\hline
\end{tabular}
\begin{list}{}{}
\item[$^{\mathrm{a}}$] $\Omega_0$: solid-body rotation rate; $\cs$: effective sound speed; $M_0$: initial envelope mass; $\tacc$: accretion time; $\tau_\el{ads}$: adsorption timescale for \w{} at the edge of the initial cloud; $M_\el{d}$: disk mass at $\tacc$.
\item[$^{\mathrm{b}}$] Case 3 is our standard parameter set and Case 7 is our reference set.
\end{list}
\end{table}

\tb{pgrid} also lists the accretion time and the adsorption timescale for \w{} at the edge of the initial envelope. For comparison, \citet{evans09a} found a median timescale for the embedded phase (Class 0 and I) of \scit{5.4}{5} yr from observations. It should be noted that the end point of our model ($\tacc$) is not yet representative of a typical T Tauri disk (see Section \ref{subsec:stanrestemp}). Nevertheless, it allows an exploration of how the chemistry responds to plausible changes in the environment.


\subsection{Adsorption and desorption}
\label{subsec:adsdes}
The adsorption and desorption of CO and \w{} are solved in a Lagrangian frame. When the time-dependent density, velocity and temperature profiles have been calculated, the envelope is populated by a number of parcels of material (typically 12,000) at $t=0$. They fall in towards the star or disk according to the velocity profiles. The density and temperature along each parcel's infall trajectory are used as input to solve the adsorption-desorption balance. Both species start fully in the gas phase. The envelope is kept static for \scit{3}{5} yr before the onset of collapse to simulate the pre-stellar core phase. This is the same value as used by \citet{rodgers03a} and \citetalias{brinch08b}, and it is consistent with recent observations by \citet{enoch08a}. The amount of gaseous material left over near the end of the pre-collapse phase is also consistent with observations, which show that the onset of \w{} ice formation is around an $A_\el{V}$ of 3 \citep{whittet01a}. In six of our eight parameter sets, the adsorption timescales of \w{} at the edge of the cloud are shorter than the combined collapse and pre-collapse time (\tb{pgrid}), so all \w{} is expected to freeze out before entering the disk. Because of the larger cloud size, the adsorption timescales are much longer in Cases 1 and 5, and some \w{} may still be in the gas phase when it reaches the disk.

No chemical reactions are included other than adsorption and thermal desorption, so the total abundance of CO and \w{} in each parcel remains constant. The adsorption rate in \pcc \ps{} is taken from \citet{charnley01a}:
\begin{equation}
\label{eq:adsrate}
R_\el{ads}(X) = (\scim{4.55}{-18}\,\el{cm}^3\,\el{K}^{-1/2}\,\el{s}^{-1})n_\el{H}n_\el{g}(X)\sqrt{\frac{T_\el{g}}{M(X)}}\,,
\end{equation}
where $n_\el{H}$ is the total hydrogen density, $T_\el{g}$ the gas temperature, $n_\el{g}(X)$ the gas-phase abundance of species $X$ and $M(X)$ its molecular weight. The numerical factor assumes unit sticking efficiency, a grain radius of 0.1 \micron{} and a grain abundance $n_\el{gr}$ of \ten{-12} with respect to H$_2$.

The thermal desorption of CO and \w{} is a zeroth-order process:
\begin{equation}
\label{eq:desrate}
R_\el{des}(X) = (\scim{1.26}{-21}\,\el{cm}^2)n_\el{H}f(X)\nu(X)\exp\left[-\frac{E_\el{b}(X)}{kT_\el{d}}\right]\,,
\end{equation}
where $T_\el{d}$ is the dust temperature and
\begin{equation}
\label{eq:covfrac}
f(X) = \min\left[1,\frac{n_\el{s}(X)}{N_\el{b}n_\el{gr}}\right]\,,
\end{equation}
with $n_\el{s}(X)$ the solid abundance of species $X$ and $N_\el{b}=10^6$ the typical number of binding sites per grain. The numerical factor in \eq{desrate} assumes the same grain properties as in \eq{adsrate}. The pre-exponential factor, $\nu(X)$, and the binding energy, $E_\el{b}(X)/k$, are set to \scit{7}{26} cm$^{-2}$ \ps{} and 855 K for CO and to \scit{1}{30} cm$^{-2}$ \ps{} and 5773 K for \w{} \citep{bisschop06a,fraser01a}.

\begin{table}
\caption{Binding energies and desorbing fractions for the four-flavour CO evaporation model.$^{\mathrm{a}}$}
\label{tb:4flav}
\centering
\begin{tabular}{ccc}
\hline\hline
Flavour & $E_\el{b}(\el{CO})/k$ (K)$^{\mathrm{b}}$ & Fraction$^{\mathrm{c}}$ \\
\hline
1       & \phantom{0}855                  & 0.350 \\
2       & \phantom{0}960                  & 0.455 \\
3       & 3260                            & 0.130 \\
4       & 5773                            & 0.065 \\
\hline
\end{tabular}
\begin{list}{}{}
\item[$^{\mathrm{a}}$] Based on \citet{viti04a}.
\item[$^{\mathrm{b}}$] The rates for Flavours 1--3 are computed from \eq{desrate} with $X=\el{CO}$. The rate for Flavour 4 is equal to the \w{} desorption rate.
\item[$^{\mathrm{c}}$] These numbers indicate fractions of adsorbing CO: 35\% of all adsorbing CO becomes Flavour 1, and so on.
\end{list}
\end{table}

\begin{figure*}
\resizebox{\hsize}{!}{\includegraphics{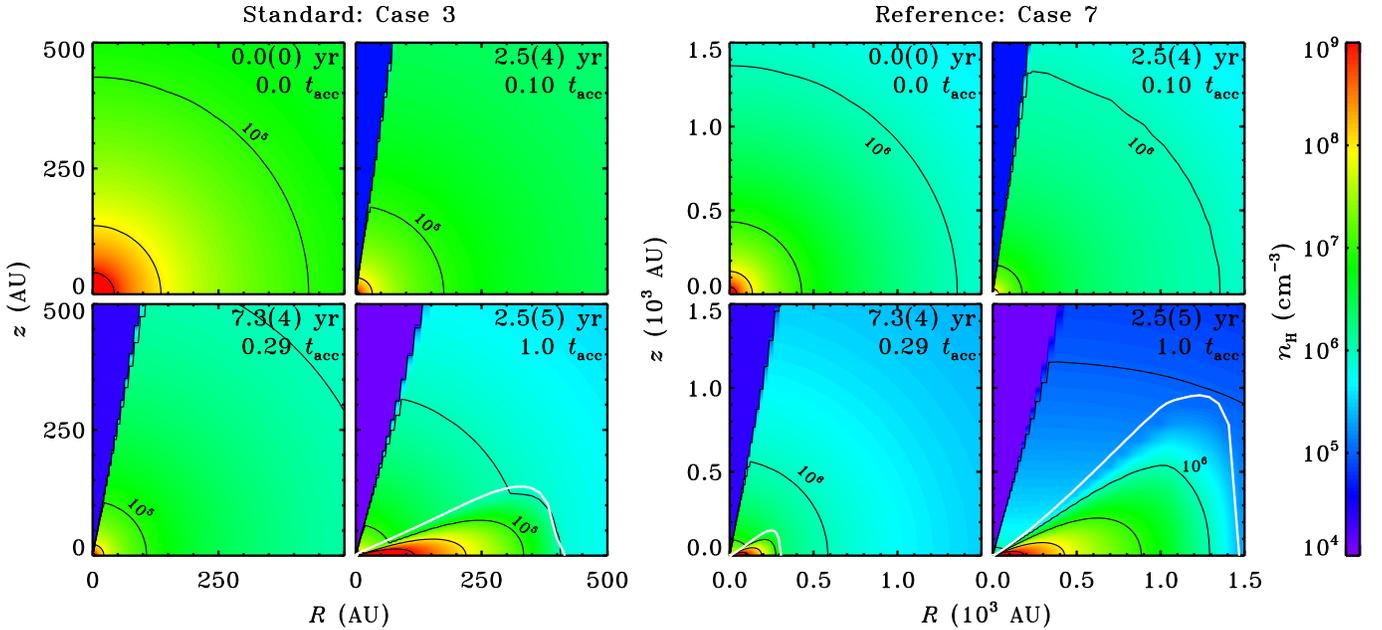}}
\caption{Total density at four time steps for our standard model (Case 3; left) and our reference model (Case 7; right). The time is given in years as well as in units of the accretion time, with $a(b)$ meaning \scit{a}{b}. The density contours increase by factors of ten going inwards; the \ten{5}-\pcc{} contours are labelled in the standard panels and the \ten{6}-\pcc{} contours in the reference panels. The white curves indicate the surface of the disk as defined in Section \ref{subsec:disk} (only visible in three panels). Note the different scale between the two sets of panels.}
\label{fig:standen}
\end{figure*}

Using a single $E_\el{b}(\el{CO})$ value means that all CO evaporates at the same temperature. This would be appropriate for a pure CO ice, but not for a mixed CO-\w{} ice as is likely to form in reality. During the warm-up phase, part of the CO is trapped inside the \w{} ice until the temperature becomes high enough for the \w{} to evaporate. Recent laboratory experiments suggest that CO desorbs from a CO-\w{} ice in four steps \citep{collings04a}. This can be simulated with four ``flavours'' of CO ice, each with a different $E_\el{b}(\el{CO})$ value \citep{viti04a}. For each flavour, the desorption is assumed to be zeroth order. The four-flavour model is summarised in \tb{4flav}.


\section{Results}
\label{sec:stanres}
Results are presented in this section for our standard and reference models (Cases 3 and 7) as described in Section \ref{subsec:params}. These cases will be compared to the other parameter sets in Section \ref{subsec:paramgrid}. Appendix \ref{sec:appa} describes a formula to estimate the disk formation efficiency, defined as $M_\el{d}/M_0$ at the end of the collapse phase, based on a fit to our model results.


\subsection{Density profiles and infall trajectories}
\label{subsec:stanresphys}
In our standard model (Case 3), the envelope collapses in \scit{2.5}{5} yr to give a star of 0.94 $M_\odot$ and a disk of 0.05 $M_\odot$. The remaining 0.01 $M_\odot$ has disappeared through the bipolar outflow. The centrifugal radius in our standard model at $\tacc$ is 4.9 AU, but the disk has spread to 400 AU at that time due to angular momentum redistribution. The densities in the disk are high: more than \ten{9} \pcc{} at the midplane inside of 120 AU (\fig{standen}, left) and more than \ten{14} \pcc{} near 0.3 AU. The corresponding surface densities of the disk are 2.0 g cm$^{-2}$ at 120 AU and 660 g cm$^{-2}$ at 0.3 AU.

Due to the higher rotation rate, our reference model (Case 7) gets a much higher disk mass: 0.43 $M_\odot$. This value is consistent with the mass of 0.4 $M_\odot$ reported by \citetalias{brinch08b}. Overall, the reference densities from our semi-analytical model (\fig{standen}, right) compare well with those from their more realistic hydrodynamical simulations; the differences are generally less than a factor of two.

In both cases, the disk first emerges at \scit{2}{4} yr, when the FHC contracts to become the protostar, but it is not until a few \ten{4} yr later that the disk becomes visible on the scale of \fig{standen}. The regions of high density ($n_\el{H}>10^5$--$10^6$ \pcc) are still contracting at that time, but the growing disks eventually cause them to expand again.

\begin{figure*}
\centering
\includegraphics[width=17cm]{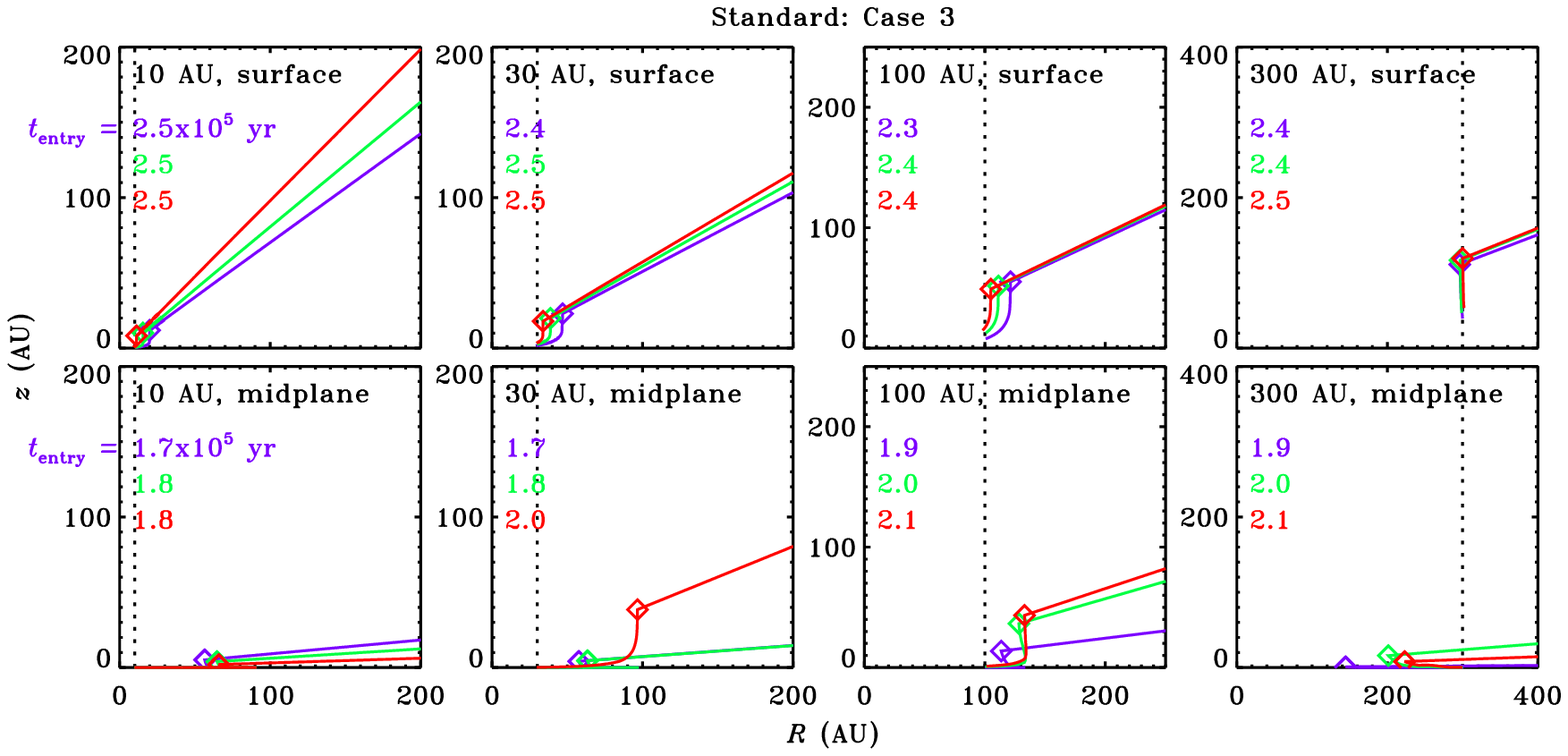}
\caption{Infall trajectories for parcels in our standard model (Case 3) ending up near the surface (top panels) or at the midplane (bottom panels) at radial positions of 10, 30, 100 and 300 AU (dotted lines) at $t=\tacc$. Each panel contains trajectories for three parcels, which are illustrative for material ending up at the given location. Trajectories are only drawn up to $t=\tacc$. Diamonds indicate where each parcel enters the disk; the time of entry is given in units of \ten{5} yr. Note the different scales between some panels.}
\label{fig:stantraj}
\end{figure*}

\begin{figure*}
\centering
\includegraphics[width=17cm]{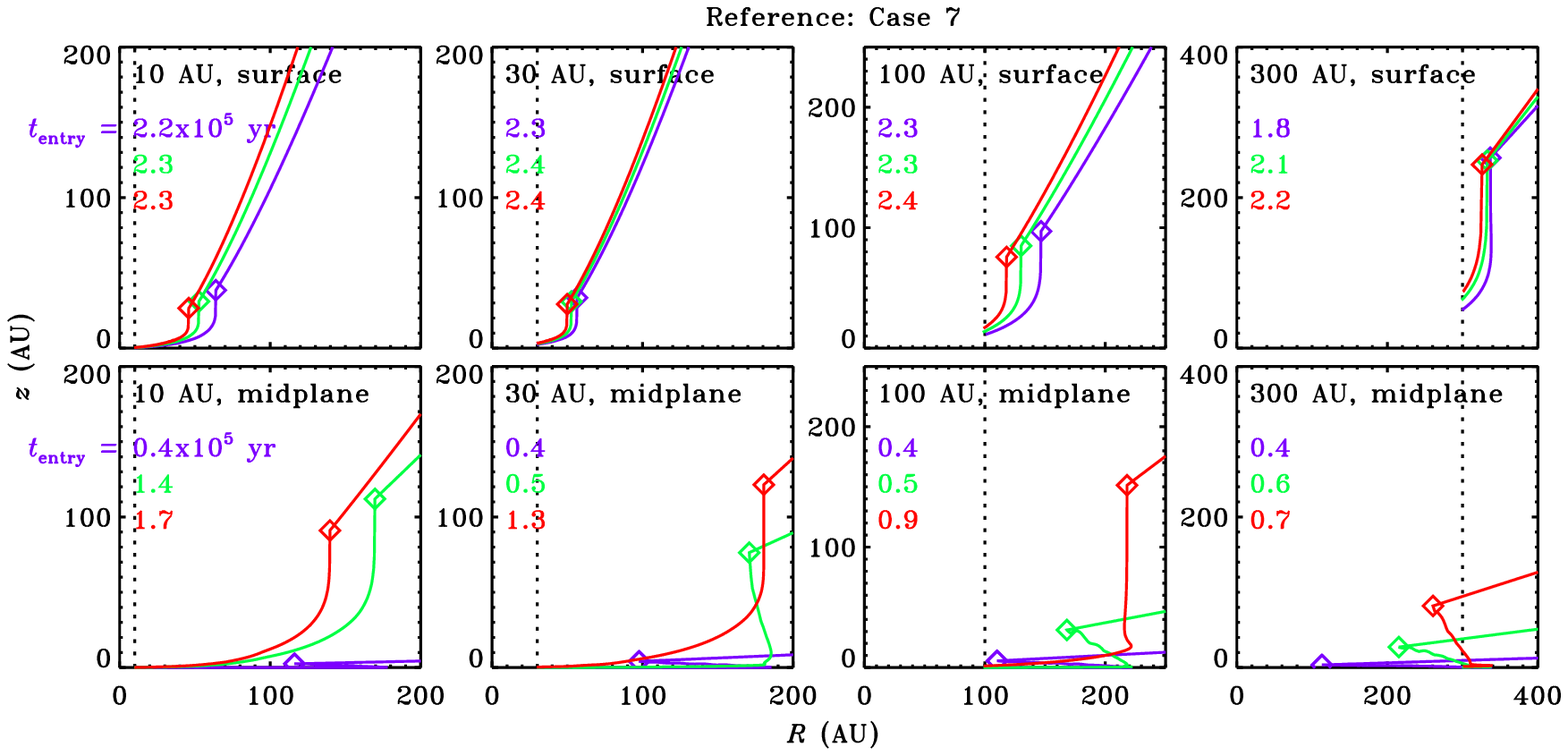}
\caption{Same as \fig{stantraj}, but for our reference model (Case 7).}
\label{fig:reftraj}
\end{figure*}

Material falls in along nearly radial streamlines far out in the envelope and deflects towards the midplane closer in. When a parcel enters the disk, it follows the radial motion caused by the viscous evolution and accretion of more material from the envelope. At any time, conservation of angular momentum causes part of the disk to move inwards and part of it to move outwards. An individual parcel entering the disk may move out for some time before going further in. This leads the parcel through several density and temperature regimes, which may affect the gas-ice ratios or the chemistry in general. The back-and-forth motion occurs especially at early times, when the entire system changes more rapidly than at later times. The parcel motions are visualised in Figs.\ \ref{fig:stantraj} and \ref{fig:reftraj}, where infall trajectories are drawn for 24 parcels ending up at one of eight positions at $\tacc$: at the midplane or near the surface at radial distances of 10, 30, 100 and 300 AU. Only parcels entering the disk before $t\approx\scim{2}{5}$ yr in our standard model or $t\approx\scim{1}{5}$ yr in our reference model undergo the back-and-forth motion. The parcels ending up near the midplane all entered the disk earlier than the ones ending up at the surface.

Accretion from the envelope onto the disk occurs in an inside-out fashion. Because of the geometry of the disk (\fig{boundary}), most of the material enters near the outer edge and prevents the older material from moving further out. Our flow inside the disk is purely laminar, so some material near the midplane does move outwards underneath the newer material at higher altitudes.

\begin{figure*}
\resizebox{\hsize}{!}{\includegraphics{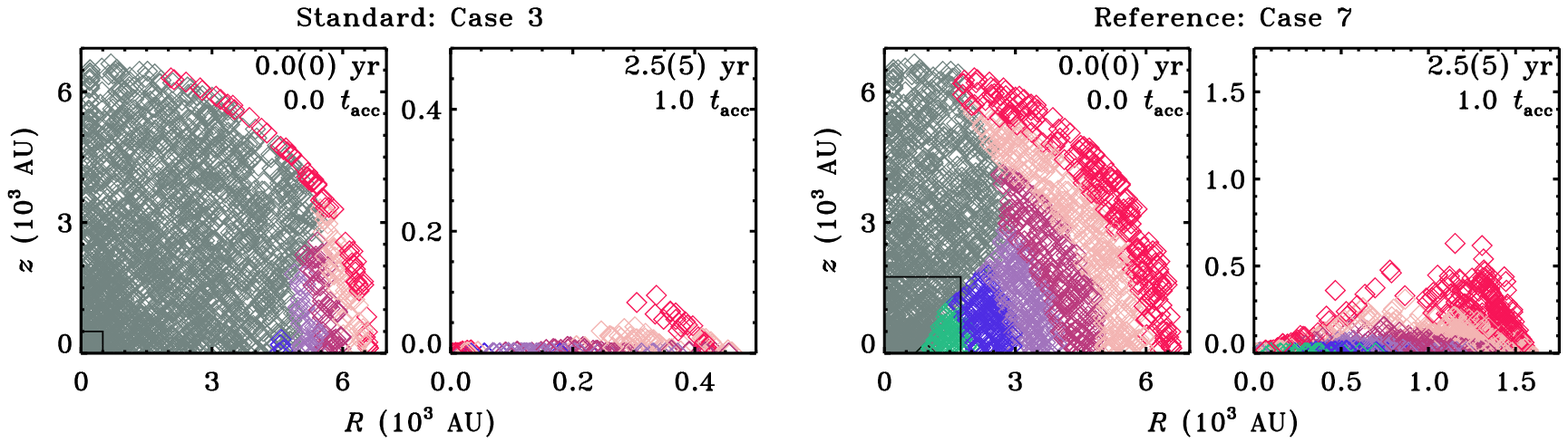}}
\caption{Position of parcels of material in our standard model (Case 3; left) and our reference model (Case 7; right) at the onset of collapse ($t=0$) and at the end of the collapse phase ($t=\tacc$). The parcels are colour-coded according to their initial position. A layered structure is visible in the disk, with material near the surface and at the outer edge originating from further out in the envelope than the material near the midplane. This effect is most pronounced in our reference model. The grey parcels from $t=0$ are in the star or have disappeared through the outflow at $t=\tacc$. Note the different spatial scale between the two panels of each set; the small box in the left panel indicates the scale of the right panel.}
\label{fig:stanparc}
\end{figure*}

\begin{figure*}
\resizebox{\hsize}{!}{\includegraphics{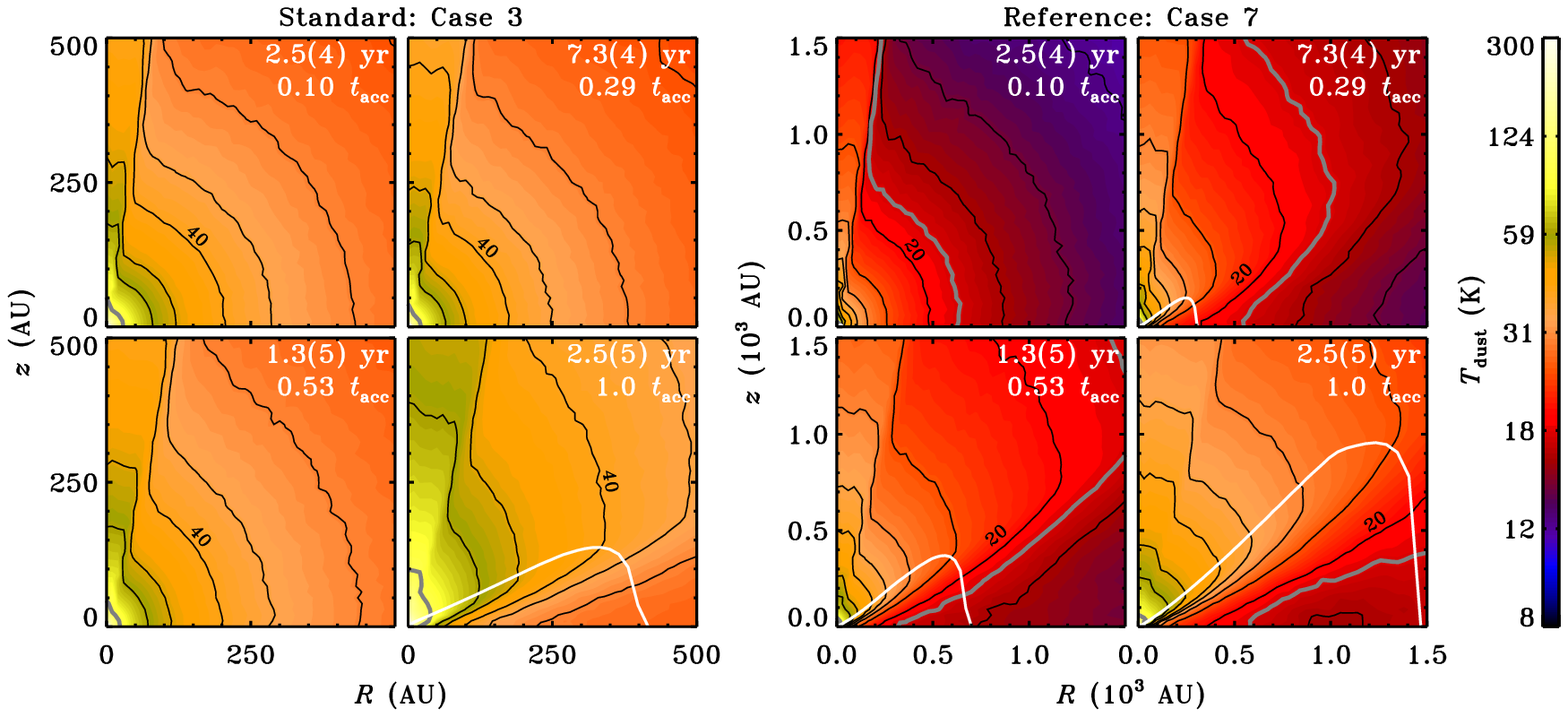}}
\caption{Dust temperature, as in \fig{standen}. Contours are drawn at 100, 60, 50, 40, 35, 30, 25, 20, 18, 16, 14 and 12 K. The 40- and 20-K contours are labelled in the standard and reference panels, respectively. The 18- and 100-K contours are drawn as thick grey lines. The white curves indicate the surface of the disk as defined in Section \ref{subsec:disk} (only visible in four panels).}
\label{fig:stantemp}
\end{figure*}

Because of the low rotation rate in our standard model, the disk does not really begin to build up until \scit{1.5}{5} yr ($0.6\tacc$) after the onset of collapse. In addition, most of the early material to reach the disk proceeds onto the star before the end of the accretion phase, so the disk at $\tacc$ consists only of material from the edge of the original cloud (\fig{stanparc}, left two panels).

The disk in our reference model, however, begins to form right after the FHC--protostar transition at \scit{2}{4} yr. As in the standard model, a layered structure is visible in the disk, but it is more pronounced here. At the end of the collapse, the midplane consists mostly of material that was originally close to the centre of the envelope (\fig{stanparc}, right two panels). The surface and outer parts of the disk are made up primarily of material from the outer parts of the envelope. This was also reported by \citetalias{brinch08b}.


\subsection{Temperature profiles}
\label{subsec:stanrestemp}
When the star turns on at \scit{2}{4} yr, the envelope quickly heats up and reaches more than 100 K inside of 10 AU. As the disk grows, its interior is shielded from direct irradiation and the midplane cools down again. At the same time, the remnant envelope material above the disk becomes less dense and warmer. As in \citet{whitney03a}, the outflow has some effect on the temperature profile. Photons emitted into the outflow can scatter and illuminate the disk from the top, causing a higher disk temperature beyond $R\approx200$ AU than if there were no outflow cone. At smaller radii, the disk temperature is lower than in a no-outflow model. Without the outflow, the radiation would be trapped in the inner envelope and inner disk, increasing the temperature at small radii.

At $t=\tacc$ in our standard model, the 100- and 18-K isotherms (where \w{} and pure CO evaporate) intersect the midplane at 20 and 2000 AU (\fig{stantemp}, left). The disk in our reference model is denser and therefore colder: it reaches 100 and 18 K at 5 and 580 AU on the midplane (\fig{stantemp}, right). Our radiative transfer method is a more rigorous way to obtain the dust temperature than the diffusion approximation used by \citetalias{brinch08b}, so our temperature profiles are more realistic than theirs.

Compared to typical T Tauri disk models \citep[e.g.][]{dalessio98a,dalessio99a,dalessio01a}, our standard disk at $\tacc$ is warmer. It is 81 K at 30 AU on the midplane, while the closest model from the D'Alessio catalogue is 28 K at that point. If our model is allowed to run beyond $\tacc$, part of the disk accretes further onto the star. At $t=4\tacc$ (\ten{6} yr), the disk mass goes down to 0.03 $M_\odot$. The luminosity of the star decreases during this period \citep{dantona94a}, so the disk cools down: the midplane temperature at 30 AU is now 42 K. Meanwhile, the dust is likely to grow to larger sizes, which would further decrease the temperatures \citep{dalessio01a}. Hence, it is important to realise that the normal end point of our models does not represent a ``mature'' T Tauri star and disk as typically discussed in the literature.

\begin{figure}
\resizebox{\hsize}{!}{\includegraphics{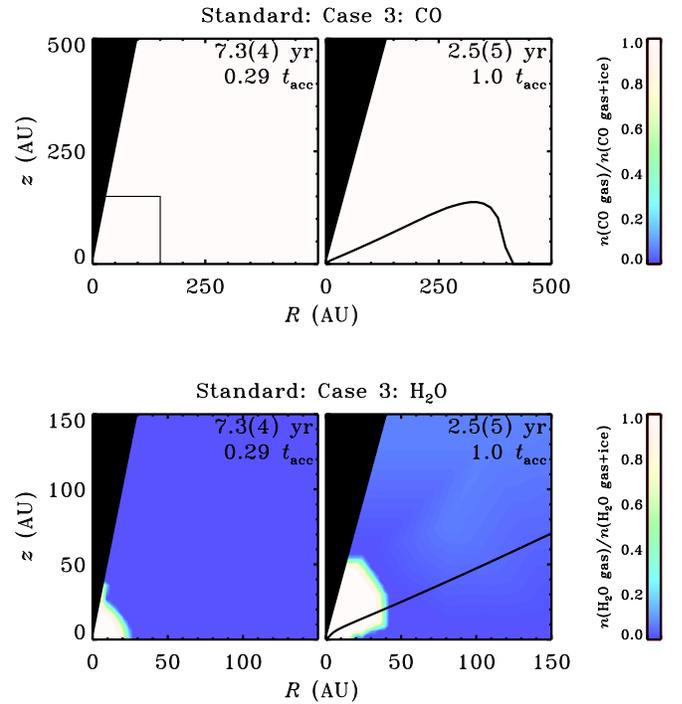}}
\caption{Gaseous CO as a fraction of the total CO abundance (top) and idem for \w{} (bottom) at two time steps for our standard model (Case 3). The black curves indicate the surface of the disk (only visible in two panels). The black area near the pole is the outflow, where no abundances are computed. Note the different spatial scale between the two panels of each set; the small box in the left CO panel indicates the scale of the \w{} panels.}
\label{fig:stancoh2o}
\end{figure}

\begin{figure}
\resizebox{\hsize}{!}{\includegraphics{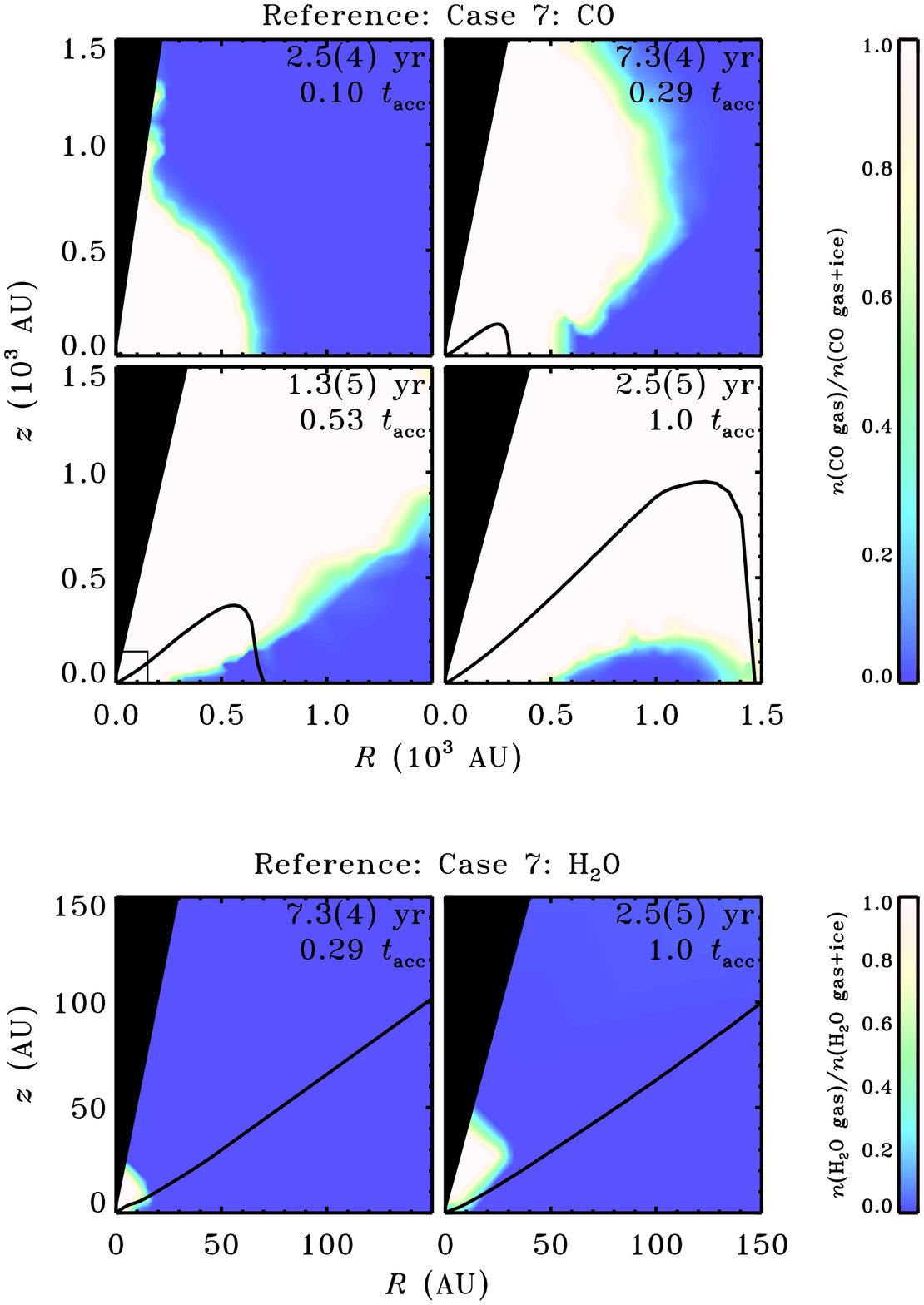}}
\caption{Same as \fig{stancoh2o}, but for our reference model (Case 7). The CO gas fraction is plotted on a larger scale and at two additional time steps.}
\label{fig:refcoh2o}
\end{figure}


\subsection{Gas and ice abundances}
\label{subsec:stanreschem}
Our two species, CO and \w, begin entirely in the gas phase. They freeze out during the static pre-stellar core phase from the centre outwards due to the density dependence of \eq{adsrate}. After the pre-collapse phase of \scit{3}{5} yr, only a few tenths of per cent of each species is still in the gas phase at 3000 AU. About 30\% remains in the gas phase at the edge of the envelope.

Up to the point where the disk becomes important and the system loses its spherical symmetry, our model gives the same results as the one-dimensional collapse models: the temperature quickly rises to a few tens of K in the collapsing region, driving some CO (evaporating around 18 K in the one-flavour model) into the gas phase, but keeping \w{} (evaporating around 100 K) on the grains.

As the disk grows in mass, it provides an increasingly large body of material that is shielded from the star's radiation, and that is thus much colder than the surrounding envelope. However, the disk in our standard model never gets below 18 K before the end of the collapse (Section \ref{subsec:stanrestemp}), so CO remains in the gas phase (\fig{stancoh2o}, top). Note that trapping of CO in the \w{} ice is not taken into account here; this possibility will be discussed in Section \ref{subsec:stanflav}.

The disk in our reference model is more massive and therefore colder. After about \scit{5}{4} yr, the outer part drops below 18 K. CO arriving in this region re-adsorbs onto the grains (\fig{refcoh2o}, top). Another \scit{2}{5} yr later, at $t=\tacc$, 19\% of all CO in the disk is in solid form. Moving out from the star, the first CO ice is found at the midplane at 400 AU. The solid fraction gradually increases to unity at 600 AU. At $R=1000$ AU, nearly all CO is solid up to an altitude of 170 AU. The solid and gaseous CO regions meet close to the 18-K surface. The densities throughout most of the disk are high enough that once a parcel of material goes below the CO desorption temperature, all CO rapidly disappears from the gas. The exception to this rule occurs at the outer edge, near 1500 AU, where the adsorption and desorption timescales are longer than the dynamical timescales of the infalling material. Small differences between the trajectories of individual parcels then cause some irregularities in the gas-ice profile.

The region containing gaseous \w{} is small at all times during the collapse. At $t=\tacc$, the snow line (the transition of \w{} from gas to ice) lies at 15 AU at the midplane in our standard model (\fig{stancoh2o}, bottom). The surface of the disk holds gaseous \w{} out to $R=41$ AU, and overall 13\% of all \w{} in the disk is in the gas phase. This number is much lower in the colder disk of our reference model: only 0.4\%. The snow line now lies at 7 AU and gaseous \w{} can be found out to 17 AU in the disk's surface layers (\fig{refcoh2o}, bottom).

Using the adsorption-desorption history of all the individual infalling parcels, the original envelope can be divided into several chemical zones. This is trivial for our standard model. All CO in the disk is in the gas phase and it has the same qualitative history: it freezes out before the onset of collapse and quickly evaporates as it falls in. \w{} also freezes out initially and only returns to the gas phase if it reaches the inner disk.

\begin{figure}
\resizebox{\hsize}{!}{\includegraphics{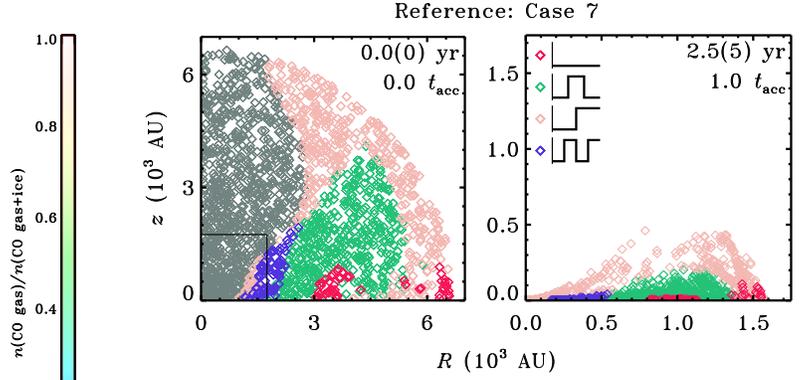}}
\caption{Same as \fig{stanparc}, but only for our reference model (Case 7) and with a different colour scheme to denote the CO adsorption-desorption behaviour. In all parcels, CO adsorbs during the pre-collapse phase. Red parcels: CO remains adsorbed; green parcels: CO desorbs and re-adsorbs; pink parcels: CO desorbs and remains desorbed; blue parcels: CO desorbs, re-adsorbs and desorbs once more. The fraction of gaseous CO in each type of parcel as a function of time is indicated schematically in the inset in the right panel. The grey parcels from $t=0$ are in the star or have disappeared through the outflow at $t=\tacc$. In our standard model (Case 3), all CO in the disk at $\tacc$ is in the gas phase and it all has the same qualitative adsorption-desorption history, equivalent to the pink parcels.}
\label{fig:refchemhist}
\end{figure}

Our reference model has the same general \w{} adsorption-desorption history, but it shows more variation for CO, as illustrated in \fig{refchemhist}. For the red parcels in that figure, more than half of the CO always remains on the grains after the initial freeze-out phase. On the other hand, more than half of the CO comes off the grains during the collapse for the green parcels, but it freezes out again inside the disk. The pink parcels, ending up in the inner disk or in the upper layers, remain warm enough to keep CO off the grains once it first evaporates. The blue parcels follow a more erratic temperature profile, with CO evaporating, re-adsorbing and evaporating a second time. This is related to the back-and-forth motion of some material in the disk (\fig{reftraj}).

\begin{figure*}
\centering
\includegraphics[width=17cm]{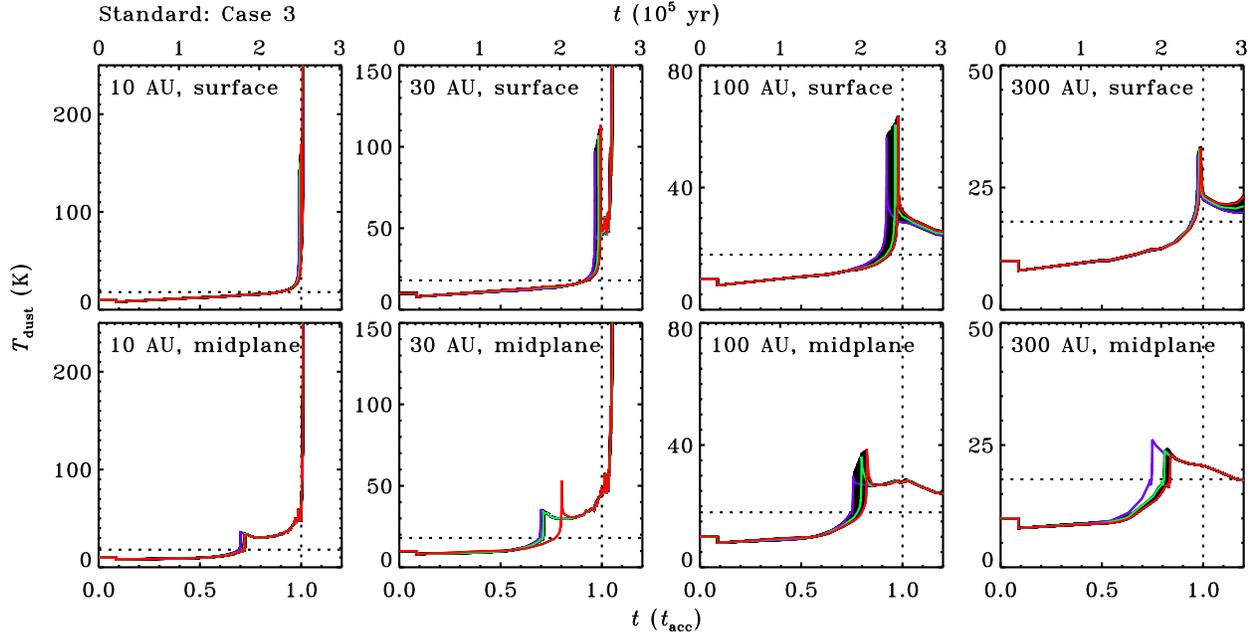}
\caption{Temperature history for parcels in our standard model (Case 3) ending up near the surface (top panels) or at the midplane (bottom panels) at radial positions of 10, 30, 100 and 300 AU at $t=\tacc$. Each panel contains between 22 and 90 curves; the coloured curves correspond to the parcels from \fig{stantraj}. The first peak in each curve occurs just before that parcel enters the disk. The dotted lines are drawn at $T_\el{d}=18$ K and $t=\tacc$. Note the different vertical scales between some panels.}
\label{fig:stanhist}
\end{figure*}

\onlfig{13}{
\begin{figure*}
\centering
\includegraphics[width=17cm]{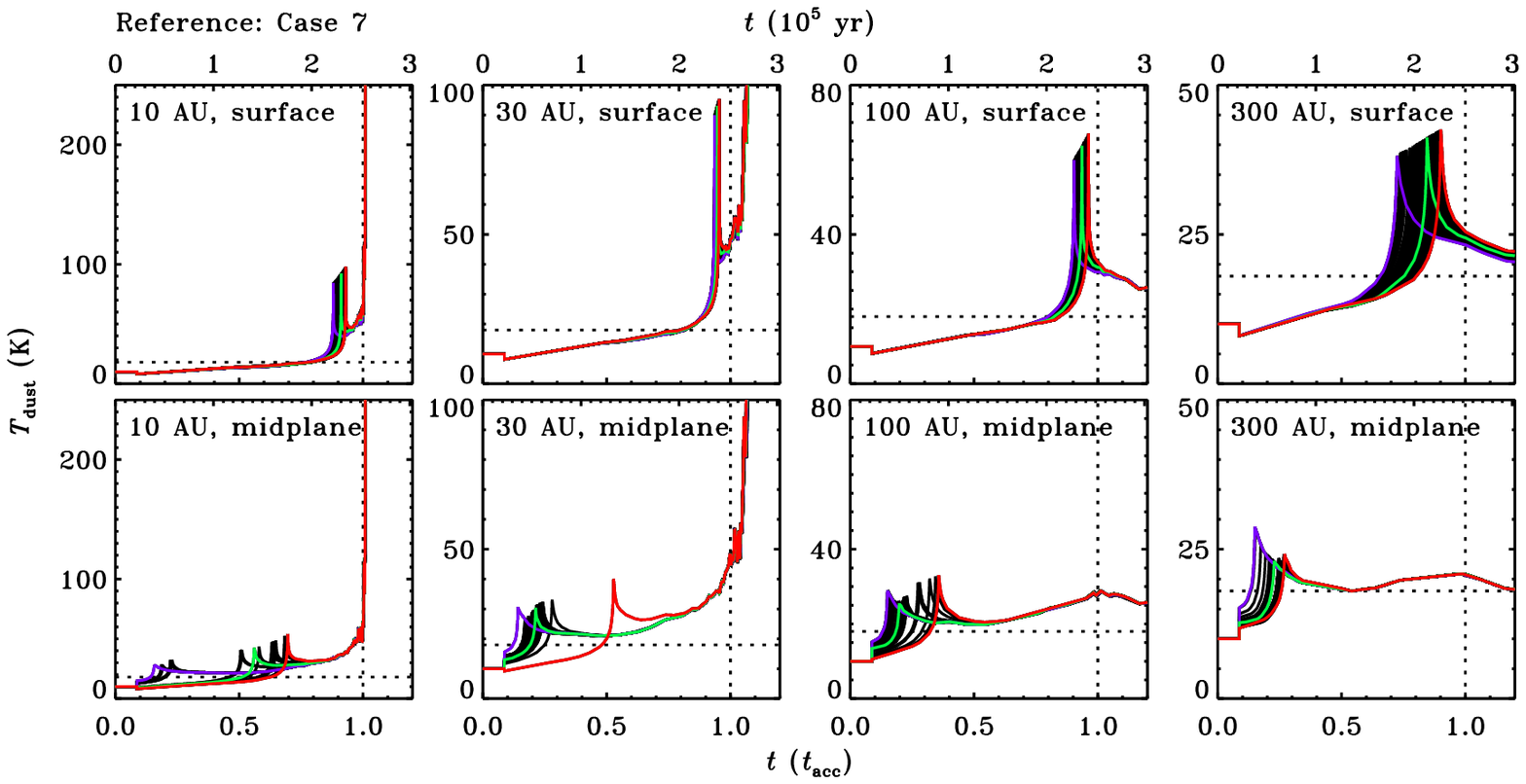}
\caption{Same as \fig{stanhist}, but for our reference model (Case 7). The coloured curves correspond to the parcels from \fig{reftraj}.}
\label{fig:refhist}
\end{figure*}
}


\subsection{Temperature histories}
\label{subsec:stantraj}
The proximity of the CO and \w{} gas-ice boundaries to the 18- and 100-K surfaces indicates that the temperature is primarily responsible for the adsorption and desorption. At $n_\el{H}=10^6$ \pcc, adsorption and desorption of CO are equally fast at $T_\el{d}=18$ K (a timescale of \scit{9}{3} yr). For a density a thousand times higher or lower, the dust temperature only has to increase or decrease by 2--3 K to maintain $k_\el{ads}=k_\el{des}$.

The exponential temperature dependence in the desorption rate [\eq{desrate}] also holds for other species than CO and \w{}, as well as for the rates of some chemical reactions. Hence, it is useful to compute the temperature history for infalling parcels that occupy a certain position at $\tacc$. Figures \ref{fig:stanhist} and \ref{fig:refhist} (the latter is only available online) show these histories for material ending up at the midplane or near the surface of the disk at radial distances of 10, 30, 100 and 300 AU. Parcels ending up inside of 10 AU have a very similar temperature history as those ending up at 10 AU, except that the final temperature of the former is higher.

Each panel in Figs.\ \ref{fig:stanhist} and \ref{fig:refhist} contains the history of several dozen parcels ending up close to the desired position. The qualitative features are the same for all parcels. The temperature is low while a parcel remains far out in the envelope. As it falls in with an ever larger velocity, there is a temperature spike as it traverses the inner envelope, followed by a quick drop once it enters the disk. Inward radial motion then leads to a second temperature rise; because of the proximity to the star, this one is higher than the first increase. For most parcels in Figs.\ \ref{fig:stanhist} and \ref{fig:refhist}, the second temperature peak does not occur until long after $\tacc$. In all cases, the shock encountered upon entering the disk is weak enough that it does not heat the dust to above the temperature caused by the stellar photons (\fig{shock}).

Based on the temperature histories, the gas-ice transition at the midplane would lie inside of 10 AU for \w{} and beyond 300 AU for CO in both our models. This is indeed where they were found to be in Section \ref{subsec:stanreschem}. The transition for a species with an intermediate binding energy, such as \fmh, is then expected to be between 10 and 100 AU, if its abundance can be assumed constant throughout the collapse.

The dynamical timescales for the infalling material before it enters the disk are between \ten{4} and \ten{5} yr. The timescales decrease as it approaches the disk, due to the rapidly increasing velocities. Once inside the disk, the material slows down again and the dynamical timescales return to \ten{4}--\ten{5} yr. The adsorption timescales for CO and \w{} are initially a few \ten{5} yr, so they exceed the dynamical timescale before entering the disk. Depletion occurs nonetheless because of the pre-collapse phase with a duration of \scit{3}{5} yr. The higher densities in the disk cause the adsorption timescales to drop to 100 yr or less. If the temperature approaches (or crosses) the desorption temperature for CO or \w{}, the corresponding desorption timescale becomes even shorter than the adsorption timescale. Overall, the timescales for these specific chemical processes (adsorption and desorption) in the disk are shorter by a factor of 1000 or more than the dynamical timescales.

At some final positions, there is a wide spread in the time that the parcels spend at a given temperature. This is especially true for parcels ending up near the midplane inside of 100 AU in our reference model. All of the midplane parcels ending up near 10 AU exceed 18 K during the collapse; the first one does so at \scit{3.5}{4} yr after the onset of collapse, the last one at \scit{1.6}{5} yr. Hence, some parcels at this final position spend more than twice as long above 18 K than others. This does not appear to be relevant for the gas-ice ratio, but it is important for the formation of more complex species \citep{garrod06a}. This will be discussed in more detail in Section \ref{subsec:complex}.


\section{Discussion}
\label{sec:disc}

\subsection{Model parameters}
\label{subsec:paramgrid}
When the initial conditions of our model are modified (Section \ref{subsec:params}), the qualitative chemistry results do not change. In Cases 3, 4 and 8, the entire disk at $\tacc$ is warmer than 18 K, and it contains no solid CO\@. In the other cases, the disk provides a reservoir of relatively cold material where CO, which evaporates early on in the collapse, can return to the grains. \w{} can only desorb in the inner few AU of the disk and remnant envelope.

Figures \ref{fig:gridden} and \ref{fig:gridtemp} show the density and dust temperature at $\tacc$ for each parameter set; our standard and reference models are the top and bottom panel in the second column (Case 3 and 7). Several trends are visible:

\begin{figure*}
\centering
\includegraphics[width=17cm]{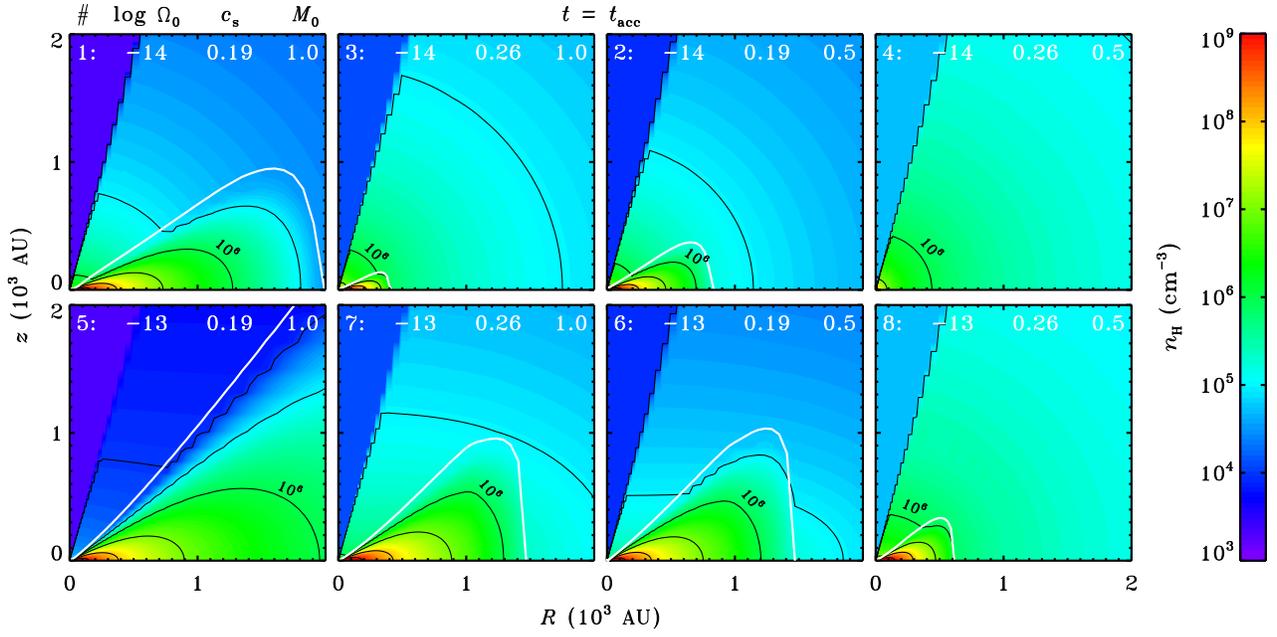}
\caption{Total density at $t=\tacc$ for each parameter set in our grid. The numbers at the top of each panel are the parameter set number, the rotation rate ($\log\Omega_0$ in \ps), the sound speed (km \ps) and the initial mass ($M_\odot$). The density contours increase by factors of ten going inwards; the \ten{6}-\pcc{} contour is labelled in each panel. The white curves indicate the surfaces of the disks; the disk for Case 4 is too small to be visible.}
\label{fig:gridden}
\end{figure*}

\begin{figure*}
\centering
\includegraphics[width=17cm]{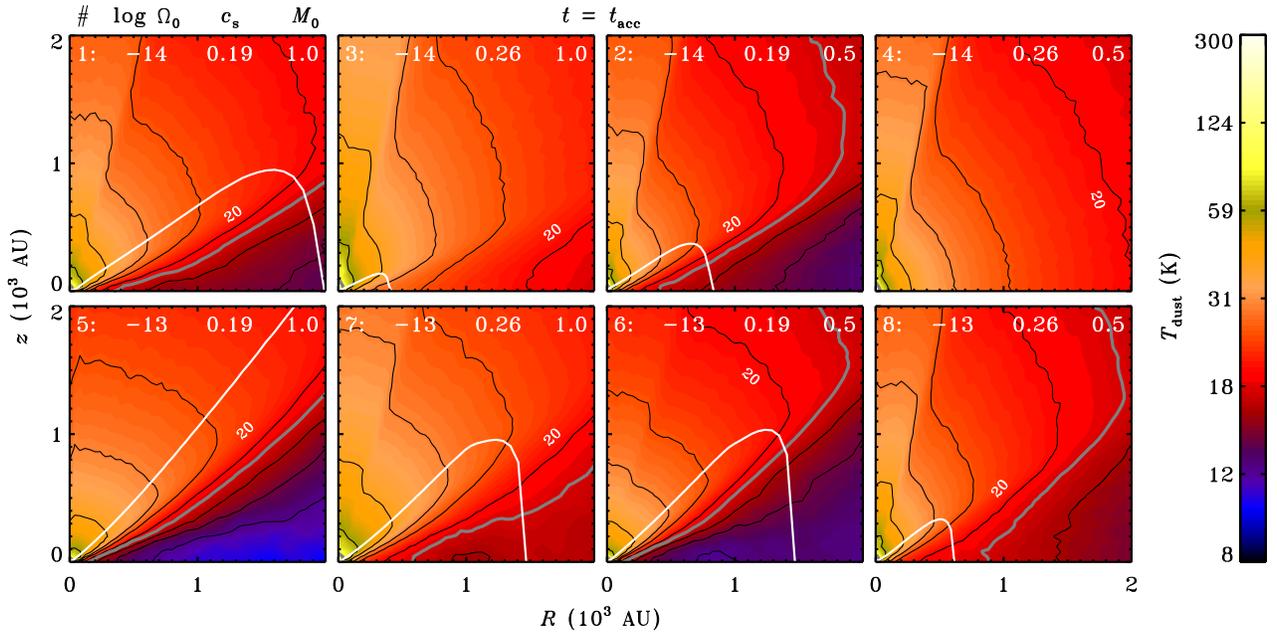}
\caption{Dust temperature, as in \fig{gridden}. The temperature contours are drawn at 100, 60, 40, 30, 25, 20, 18, 16, 14 and 12 K from the centre outwards; the 20-K contour is labelled in each panel.}
\label{fig:gridtemp}
\end{figure*}

\begin{itemize}
\item[$\bullet$] With a lower sound speed (Cases 1, 2, 5 and 6), the overall accretion rate ($\dot{M}$) is smaller so the accretion time increases ($\tacc\propto\cs^{-3}$). The disk can now grow larger and more massive. In our standard model, the disk is 0.05 $M_\odot$ at $\tacc$ and extends to about 400 AU radially. Decreasing the sound speed to 0.19 km \ps{} (Case 1) results in a disk of 0.22 $M_\odot$ and nearly 2000 AU. The lower accretion rate also reduces the stellar luminosity. These effects combine to make the disk colder in the low-$\cs$ cases.

\item[$\bullet$] With a lower rotation rate (Cases 1--4), the infall occurs in a more spherically symmetric fashion. Less material is captured in the disk, which remains smaller and less massive. From our reference to our standard model, the disk mass goes from 0.43 to 0.05 $M_\odot$ and the radius from 1400 to 400 AU. The larger accretion onto the star causes a higher luminosity. Altogether, this makes for a small, relatively warm disk in the low-$\Omega_0$ cases.

\item[$\bullet$] With a lower initial mass (Cases 2, 4, 6 and 8), there is less material to end up on the disk. The density profile is independent of the mass in a Shu-type collapse [\eq{sisrho}], so the initial mass is lowered by taking a smaller envelope radius. The material from the outer parts of the envelope is the last to accrete and is more likely, therefore, to end up in the disk. If the initial mass is halved relative to our standard model (as in Case 4), the resulting disk is only 0.001 $M_\odot$ and 1 AU. Our reference disk goes from 0.43 $M_\odot$ and 1400 AU to 0.16 $M_\odot$ and 600 AU (Cases 7 and 8). The luminosity at $\tacc$ is lower in the high-$M_0$ cases and the cold part of the disk ($T_\el{d}<18$ K) has a somewhat larger relative size.
\end{itemize}

\citet{dullemond06a} noted that accretion occurs closer to the star for a slowly rotating cloud than for a fast rotating cloud, resulting in a larger fraction of crystalline dust in the former case. The same effect is seen here, but overall the accretion takes place further from the star than in \citet{dullemond06a}. This is due to our taking into account the vertical structure of the disk. Our gaseous fractions in the low-$\Omega_0$ disks are higher than in the high-$\Omega_0$ disks (consistent with a higher crystalline fraction), but not because material enters the disk closer to the star. Rather, as mentioned above, the larger gas content comes from the higher temperatures throughout the disk.

Combining the density and the temperature, the fractions of cold ($T_\el{d}<18$ K), warm ($T_\el{d}>18$ K) and hot ($T_\el{d}>100$ K) material in the disk can be computed. The warm and hot fractions are listed in \tb{gridres} along with the fractions of gaseous CO and \w{} in the disk at $\tacc$. Across the parameter grid, 23--100\% of the CO is in the gas, along with 0.3--100\% of the \w\@. This includes Case 4, which only has a disk of 0.0014 $M_\odot$. If that one is omitted, at most 13\% of the \w{} in the disk at $\tacc$ is in the gas. The gaseous \w{} fractions for Cases 1, 2, 6, 7 and 8 (at most a few per cent) are quite uncertain, because the model does not have sufficient resolution in the inner disk to resolve these small amounts. These fractions may be lower by up to a factor of 10 or higher by up to a factor of 3.

\begin{table}
\caption{Summary of properties at $t=\tacc$ for our parameter grid.$^{\mathrm{a}}$}
\label{tb:gridres}
\centering
\begin{tabular}{cccccc}
\hline\hline
Case & $M_\el{d}/M{}^{\el{b}}$ & $f_\el{warm}{}^{\el{c}}$ & $f_\el{hot}{}^{\el{c}}$ & $f_\el{gas}$(CO)$^{\el{d}}$ & $f_\el{gas}$(\w)$^{\el{d}}$ \\
\hline
1 \phantom{(std)} & 0.22\phantom{0} & 0.69 & 0.004\phantom{0} & 0.62 & 0.028 \\
2 \phantom{(std)} & 0.15\phantom{0} & 0.94 & 0.035\phantom{0} & 0.93 & 0.020 \\
3 (std)           & 0.05\phantom{0} & 1.00 & 0.17\phantom{00} & 1.00 & 0.13\phantom{0} \\
4 \phantom{(std)} & 0.003           & 1.00 & 1.00\phantom{00} & 1.00 & 1.00\phantom{0} \\
5 \phantom{(std)} & 0.59\phantom{0} & 0.15 & 0.0001           & 0.23 & 0.11\phantom{0} \\
6 \phantom{(std)} & 0.50\phantom{0} & 0.34 & 0.0004           & 0.27 & 0.02\phantom{0} \\
7 (ref)           & 0.43\phantom{0} & 0.83 & 0.003\phantom{0} & 0.81 & 0.004 \\
8 \phantom{(std)} & 0.33\phantom{0} & 1.00 & 0.028\phantom{0} & 1.00 & 0.003 \\
\hline
\end{tabular}
\begin{list}{}{}
\item[$^{\mathrm{a}}$] These results are for the one-flavour CO desorption model.
\item[$^{\mathrm{b}}$] The fraction of the disk mass with respect to the total accreted mass ($M=M_\ast+M_\el{d}$).
\item[$^{\mathrm{c}}$] The fractions of warm ($T_\el{d}>18$ K) and hot ($T_\el{d}>100$ K) material with respect to the entire disk. The warm fraction also includes material above 100 K.
\item[$^{\mathrm{d}}$] The fractions of gaseous CO and \w{} with respect to the total amounts of CO and \w{} in the disk.
\end{list}
\end{table}

There is good agreement between the fractions of warm material and gaseous CO\@. In Case 5, about a third of the CO gas at $\tacc$ is gas left over from the initial conditions, due to the long adsorption timescale for the outer part of the cloud. This is also the case for the majority of the gaseous \w{} in Cases 1, 5 and 6. For the other parameter sets, $f_\el{hot}$ and $f_\el{gas}$(\w) are the same within the error margins. Overall, the results from the parameter grid show once again that the adsorption-desorption balance is primarily determined by the temperature, and that the adsorption-desorption timescales are usually shorter than the dynamical timescales.

By comparing the fraction of gaseous material at the end of the collapse to the fraction of material above the desorption temperature, the history of the material is disregarded. For example, some of the cold material was heated above 18 K during the collapse, and CO desorbed before re-adsorbing inside the disk. This may affect the CO abundance if the model is expanded to include a full chemical network. In that case, the results from \tb{gridres} only remain valid if the CO abundance is mostly constant throughout the collapse. The same caveat holds for \w\@.


\subsection{Complex organic molecules}
\label{subsec:complex}
A full chemical network is required to analyse the gas and ice abundances of more complex species. While this will be a topic for a future paper, the current CO and \w{} results, combined with recent other work, can already provide some insight.

In general, the formation of organic species can be divided into two categories: first-generation species that are formed on and reside in the grain surfaces, and second-generation species that are formed in the warm gas phase when the first-generation species have evaporated. Small first-generation species (like \meoh{}) are efficiently formed during the pre-collapse phase \citep{garrod06a}. Their gas-ice ratios should be similar to that of \w{}, due to the similar binding energies.

Larger first-generation species such as methyl formate (\mf) can be formed on the grains if material spends at least several \ten{4} yr at 20--40 K. The radicals involved in the surface formation of \mf{} (HCO and CH$_3$O) are not mobile enough at lower temperatures and are not formed efficiently enough at higher temperatures. A low surface abundance of CO (at temperatures above 18 K) does not hinder the formation of \mf: HCO and CH$_3$O are formed from reactions of OH and H with \fmh{}, which is already formed at an earlier stage and which does not evaporate until $\sim$40 K \citep{garrod06a}. Cosmic-ray-induced photons are available to form OH from \w{} even in the densest parts of the disk and envelope \citep{shen04a}.

As shown in Section \ref{subsec:stantraj}, many of the parcels ending up near the midplane inside of $\sim$300 AU in our standard model spend sufficient time in the required temperature regime to allow for efficient formation of \mf{} and other complex organics. Once formed, these species are likely to assume the same gas-ice ratios as \w{} and the smaller organics. They evaporate in the inner 10--20 AU, so in the absence of mixing, complex organics would only be observable in the gas phase close to the star. The Atacama Large Millimeter/submillimeter Array (ALMA), currently under construction, will be able to test this hypothesis.

The gas-phase route towards complex organics involves the hot inner envelope ($T_\el{d}>100$ K), also called the hot core or hot corino in the case of low-mass protostars \citep{ceccarelli04a,bottinelli04a,bottinelli07a}. Most of the ice evaporates here and a rich chemistry can take place if material spends at least several \ten{3} yr in the hot core \citep{charnley92a}. However, the material in the hot inner envelope in our model is essentially in freefall towards the star or the inner disk, and its transit time of a few 100 yr is too short for complex organics to be formed abundantly \citep[see also][]{schoier02a}. Additionally, the total mass in this region is very small: about a per cent of the disk mass. In order to explain the observations of second-generation complex molecules, there has to be some mechanism to keep the material in the hot core for a longer time. Alternatively, it has recently been suggested that molecules typically associated with hot cores may in fact form on the grain surfaces as well \citep{garrod08a}.


\subsection{Mixed CO-\w{} ices}
\label{subsec:stanflav}
In the results presented in Section \ref{sec:stanres}, all CO was taken to desorb at a single temperature. In a more realistic approach, some of it would be trapped in the \w{} ice and desorb at higher temperatures. This was simulated with four ``flavours'' of CO ice, as summarised in \tb{4flav}. With our four-flavour model, the global gas-ice profiles are largely unchanged. All CO is frozen out in the sub-18 K regions and it fully desorbs when the temperature goes above 100 K. Some 10 to 20\% remains in the solid phase in areas of intermediate temperature. In our standard model, the four-flavour variety has 15\% of all CO in the disk at $\tacc$ on the grains, compared to 0\% in the one-flavour variety. In our reference model, the solid fraction increases from 19 to 33\%.

The grain-surface formation of \fmh, \meoh, \mf{} and other organics should not be very sensitive to these variations. \fmh{} and \meoh{} are already formed abundantly before the onset of collapse, when the one- and four-flavour models predict equal amounts of solid CO\@. \fmh{} is then available to form \mf{} (via the intermediates HCO and CH$_3$O) during the collapse. The higher abundance of solid CO at 20--40 K in the four-flavour model could slow down the formation of \mf{} somewhat, because CO destroys the OH needed to form HCO \citep{garrod06a}. \fmh{} evaporates around 40 K, so \mf{} cannot be formed efficiently anymore above that temperature. On the other hand, if a multiple-flavour approach is also employed for \fmh{}, some of it remains solid above 40 K, and \mf{} can continue to be produced. Overall, then, the multiple-flavour desorption model is not expected to cause large variations in the abundances of these organic species compared to the one-flavour model.


\subsection{Implications for comets}
\label{subsec:comets}
Comets in our solar system are known to be abundant in CO and they are believed to have formed between 5 and 30 AU in the circumsolar disk \citep{bockelee04a,kobayashi07a}. However, the dust temperature in this region at the end of the collapse is much higher than 18 K for all of our parameter sets. This raises the question of how solid CO can be present in the comet-forming zone.

One possible answer lies in the fact that even at $t=\tacc$, our objects are still very young. As noted in Section \ref{subsec:stanrestemp}, the disks will cool down as they continue to evolve towards ``mature'' T Tauri systems. Given the right set of initial conditions, this may bring the temperature below 18 K inside of 30 AU. However, there are many T Tauri disk models in the literature where the temperature at those radii remains well above the CO evaporation temperature \citep[e.g.][]{dalessio98a,dalessio01a}. Specifically, models of the minimum-mass solar nebula (MMSN) predict a dust temperature of $\sim$40 K at 30 AU \citep{lecar06a}.

A more plausible solution is to turn to mixed ices. At the temperatures computed for the comet-forming zone of the MMSN, 10--20\% of all CO may be trapped in the \w{} ice. Assuming typical CO-\w{} abundance ratios, this is entirely consistent with observed cometary abundances \citep{bockelee04a}.

Large abundance variations are possible for more complex species, due to the different densities and temperatures at various points in the comet-forming zone in our model, as well as the different density and temperature histories for material ending up at those points. This seems to be at least part of the explanation for the chemical diversity observed in comets. Our current model will be extended in a forthcoming paper to include a full gas-phase chemical network to analyse these variations and compare them against cometary abundances.

The desorption and re-adsorption of \w{} in the disk-envelope boundary shock has been suggested as a method to trap noble gases in the ice and include them in comets \citep{owen92a,owen93a}. As shown in Sections \ref{subsec:shock} and \ref{subsec:stantraj}, a number of parcels in our standard model are heated to more than 100 K just prior to entering the disk. However, these parcels end up in the disk's surface. Material that ends up at the midplane, in the comet-forming zone, never gets heated above 50 K. Vertical mixing, which is ignored in our model, may be able to bring the noble-gas-containing grains down into the comet-forming zone.

Another option is episodic accretion, resulting in temporary heating of the disk (Section \ref{subsec:lims}). In the subsequent cooling phase, noble gases may be trapped as the ices reform. The alternative of trapping the noble gases already in the pre-collapse phase is unlikely. This requires all the \w{} to start in the gas phase and then freeze out rapidly. However, in reality (contrary to what is assumed in our model) it is probably formed on the grain surfaces by hydrogenation of atomic oxygen, which would not allow for trapping of noble gases.


\subsection{Limitations of the model}
\label{subsec:lims}
The physical part of our model is known to be incomplete and this may affect the chemical results. For example, our model does not include radial and vertical mixing. \citet{semenov06a} and \citet{aikawa07a} recently showed that mixing can enhance the gas-phase CO abundance in the sub-18 K regions of the disk. Similarly, there could be more \w{} gas if mixing is included. This would increase the fractions of CO and \w{} gas listed in \tb{gridres}. The gas-phase abundances can also be enhanced by allowing for photodesorption of the ices in addition to the thermal desorption considered here \citep{shen04a,oberg07a,oberg09a}. Mixing and photodesorption can each increase the total amount of gaseous material by up to a factor of 2. The higher gas-phase fractions are mostly found in the regions where the temperature is a few degrees below the desorption temperature of CO or \w{}\@.

Accretion from the envelope onto the star and disk occurs in our model at a constant rate $\dot{M}$ until all of the envelope mass is gone. However, the lack of widespread red-shifted absorption seen in interferometric observations suggests that the infall may stop already at an earlier time \citep{jorgensen07a}. This would reduce the disk mass at $\tacc$. The size of the disk is determined by the viscous evolution, which would probably not change much. Hence, if accretion stops or slows down before $\tacc$, the disk would be less dense and therefore warmer. It would also reduce the fraction of disk material where CO never desorbed, because most of that material comes from the outer edge of the original cloud (\fig{refchemhist}). Both effects would increase the gas-ice ratios of CO and \w\@.

Our results are also modified by the likely occurence of episodic accretion \citep{kenyon95a,evans09a}. In this scenario, material accretes from the disk onto the star in short bursts, separated by intervals where the disk-to-star accretion rate is a few orders of magnitude lower. The accretion bursts cause luminosity flares, briefly heating up the disk before returning to an equilibrium temperature that is lower than in our models. This may produce a disk with a fairly large ice content for most of the time, which evaporates and re-adsorbs after each accretion episode. The consequences for complex organics and the inclusion of various species in comets are unclear.


\section{Conclusions}
\label{sec:conc}
This paper presents the first results from a two-dimensional, semi-analytical model that simulates the collapse of a molecular cloud to form a low-mass protostar and its surrounding disk. The model follows individual parcels of material from the cloud into the star or disk and also tracks their motion inside the disk. It computes the density and temperature at each point along these trajectories. The density and temperature profiles are used as input for a chemical code to calculate the gas and ice abundances for carbon monoxide (CO) and water (\w) in each parcel, which are then transformed into global gas-ice profiles. Material ending up at different points in the disk spends a different amount of time at certain temperatures. These temperature histories provide a first look at the chemistry of more complex species. The main results from this paper are as follows:

\begin{itemize}
\item[$\bullet$] Both CO and \w{} freeze out towards the centre of the cloud before the onset of collapse. As soon as the protostar turns on, a fraction of the CO rapidly evaporates, while \w{} remains on the grains. CO returns to the solid phase when it cools below 18 K inside the disk. Depending on the initial conditions, this may be in a small or a large fraction of the disk (Section \ref{subsec:stanreschem}).

\item[$\bullet$] All parcels that end up in the disk have the same qualitative temperature history (\fig{stanhist}). There is one temperature peak just before entering the disk, when material traverses the inner envelope, and a second one (higher than the first) when inward radial motion brings the parcel closer to the star. In some cases, this results in multiple desorption and adsorption events during the parcel's infall history (Section \ref{subsec:stantraj}).

\item[$\bullet$] Material that originates near the midplane of the initial envelope remains at lower temperatures than material originating from closer to the poles. As a result, the chemical content of the material from near the midplane is less strongly modified during the collapse than the content of material from other regions (\fig{refchemhist}). The outer part of the disk contains the chemically most pristine material, where at most only a small fraction of the CO ever desorbed (Section \ref{subsec:stanreschem}).

\item[$\bullet$] A higher sound speed results in a smaller and warmer disk, with larger fractions of gaseous CO and \w{} at the end of the envelope accretion. A lower rotation rate has the same effect. A higher initial mass results in a larger and colder disk, and smaller gaseous CO and \w{} fractions (Section \ref{subsec:paramgrid}).

\item[$\bullet$] The infalling material generally spends enough time in a warm zone that first-generation complex organic species can be formed abundantly on the grains (\fig{stanhist}). Large differences can occur in the density and temperature histories for material ending up at various points in the disk. These differences allow for spatial abundance variations in the complex organics across the entire disk. This appears to be at least part of the explanation for the cometary chemical diversity (Sections \ref{subsec:complex} and \ref{subsec:comets}).

\item[$\bullet$] Complex second-generation species are not formed abundantly in the warm inner envelope (the hot core or hot corino) in our model, due to the combined effects of the dynamical timescales and low mass fraction in that region (Section \ref{subsec:complex}).

\item[$\bullet$] The temperature in the disk's comet-forming zone (5--30 AU from the star) lies well above the CO desorption temperature, even if effects of grain growth and continued disk evolution are taken into account. Observed cometary CO abundances can be explained by mixed ices: at temperatures of several tens of K, as predicted for the comet-forming zone, CO can be trapped in the \w{} ice at a relative abundance of a few per cent (Section \ref{subsec:comets}).
\end{itemize}


\begin{acknowledgements}
The authors are grateful to Christian Brinch, Reinout van Weeren and Michiel Hogerheijde for stimulating discussions and easy access to their data. They acknowledge the referee, Ted Bergin, whose constructive comments helped improve the original manuscript. Astrochemistry in Leiden is supported by a Spinoza Grant from the Netherlands Organization for Scientific Research (NWO) and a NOVA grant. SDD acknowledges support by a grant from The Reseach Corporation.
\end{acknowledgements}


\bibliographystyle{aa}
\bibliography{collices}

\appendix
\section{Disk formation efficiency}
\label{sec:appa}
The results from our parameter grid can be used to derive the disk formation efficiency, $\eta_\el{df}$, as a function of the sound speed, $\cs$, the solid-body rotation rate, $\Omega_0$, and the initial cloud mass, $M_0$. This efficiency can be defined as the fraction of $M_0$ that is in the disk at the end of the collapse phase ($t=\tacc$) or as the mass ratio between the disk and the star at that time. The former is used in this Appendix.

{\setlength\arraycolsep{2pt}
In order to cover a larger range of initial conditions, the physical part of our model was run on a $9^3$ grid. The sound speed was varied from 0.15 to 0.35 km \ps, the rotation rate from \ten{-14.5} to \ten{-12.5} \ps{} and the initial cloud mass from 0.1 to 2.1 $M_\odot$. The resulting $\eta_\el{df}$ at $t=\tacc$ were fitted to
\begin{equation}
\label{eq:fit}
\eta_\el{df} = \frac{M_\el{d}}{M_0} = g_1 + g_2\left[\frac{\log(\Omega_0/\el{s}^{-1})}{-13}\right]\,
\end{equation}
with
\begin{eqnarray}
\label{eq:fitg1}
g_1 & = & k_1 + k_2\left[\frac{\log(\Omega_0/\el{s}^{-1})}{-13}\right]^{q_1} + k_3\left[\frac{\cs}{0.2\,\el{km}\,\el{s}^{-1}}\right]^{q_2} + k_4\left[\frac{M_0}{M_\odot}\right]^{q_3}\,, \nonumber\\
& & 
\end{eqnarray}
\begin{equation}
\label{eq:fitg2}
g_2 = k_5 + k_6\left[\frac{\log(\Omega_0/\el{s}^{-1})}{-13}\right] + k_7\left[\frac{\cs}{0.2\,\el{km}\,\el{s}^{-1}}\right] + k_8\left[\frac{M_0}{M_\odot}\right]\,.
\end{equation}
Equation \ref{eq:fit} can give values lower than 0 or larger than 1. In those cases, it should be interpreted as being 0 or 1.
}

The best-fit values for the coefficients $k_i$ and the exponents $q_i$ are listed in \tb{fitkq}. The absolute and relative difference between the best fit and the model data have a root mean square (rms) of 0.04 and 5\%. The largest absolute and relative difference are 0.20 and 27\%. The fit is worst for a high cloud mass, a low sound speed and an intermediate rotation rate, as well as for a low cloud mass, an intermediate to high sound speed and a high rotation rate. 

\begin{table}
\caption{Coefficients and exponents for the best fit for the disk formation efficiency.}
\label{tb:fitkq}
\centering
\begin{tabular}{c r @{.} l c r @{.} l cc}
\hline\hline
Coefficient & \multicolumn{2}{c}{Value} & Coefficient & \multicolumn{2}{c}{Value} & Exponent & Value \\
\hline
$k_1$ &  2&08  & $k_5$ & $-0$&106 & $q_1$ & 0.236 \\
$k_2$ &  0&020 & $k_6$ & $-1$&539 & $q_2$ & 0.255 \\
$k_3$ &  0&035 & $k_7$ & $-0$&470 & $q_3$ & 0.537 \\
$k_4$ &  0&914 & $k_8$ & $-0$&344 & & \\
\hline
\end{tabular}
\end{table}

\begin{figure}
\resizebox{\hsize}{!}{\includegraphics{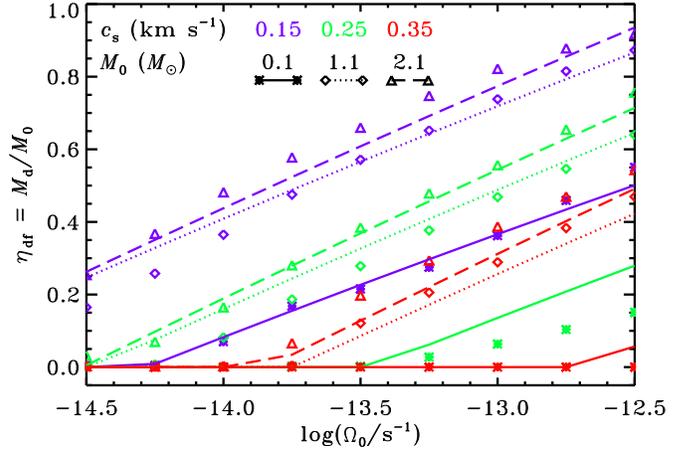}}
\caption{Disk formation efficiency as a function of the solid-body rotation rate. The model values are plotted as symbols and the fit from \eq{fit} as lines. The different values of the sound speed are indicated by colours and the different values of the initial cloud mass are indicated by symbols and line types, with the solid lines corresponding to the asterisks, the dotted lines to the diamonds and the dashed lines to the triangles.}
\label{fig:fit}
\end{figure}

\figg{fit} shows the disk formation efficiency as a function of the rotation rate, including the fit from \eq{fit}. The efficiency is roughly a quadratic function in $\log\Omega_0$, but due to the small dynamic range of this variable, the fit appears as straight lines. Furthermore, the efficiency is roughly a linear function in $\cs$ and a square root function in $M_0$.

\Online

\end{document}